%% file: paper.tex
\documentclass[sigconf]{acmart}

\usepackage[english]{babel}
\usepackage{blindtext}
\usepackage[inline]{enumitem}

\usepackage{booktabs}
\usepackage{graphicx}
\usepackage{multirow}
\usepackage{colortbl}
\usepackage{tabularx}
\usepackage{multicol}
\usepackage{balance}

\usepackage{subcaption}

\usepackage[binary-units=true]{siunitx}

\usepackage{layouts}

\acmYear{2020}\copyrightyear{2020}
\setcopyright{acmlicensed}
\acmConference[IMC '20]{ACM Internet Measurement Conference}{October 27--29, 2020}{Virtual Event, USA}
\acmBooktitle{ACM Internet Measurement Conference (IMC '20), October 27--29, 2020, Virtual Event, USA}
\acmPrice{15.00}
\acmDOI{10.1145/3419394.3423666}
\acmISBN{978-1-4503-8138-3/20/10}

\newif\ifmanualheader

\manualheadertrue

\ifmanualheader
  \makeatletter
  \g@addto@macro\@subtitlenotes{\global\let\authors=\cleanauthors}
  \makeatother
\fi

\newif\ifdraft

\drafttrue

\ifdraft
\newcommand{\todo}[1]{{\textcolor{red}{\textbf{TODO:} #1}}}

\newcommand{\md}[1]{{\color{blue}{#1}}}
\newcommand{\jp}[1]{{\color{green}{#1}}}
\newcommand{\mh}[1]{{\color{olive}{#1}}}

\else
\newcommand{\todo}[1]{}

\newcommand{\md}[1]{{}}
\newcommand{\jp}[1]{{}}
\newcommand{\mh}[1]{{}}
\fi

\newcommand{\DefineRemark}[2]{%
  \expandafter\newcommand\csname rmk-#1\endcsname{#2}%
}
\newcommand{\Remark}[1]{\csname rmk-#1\endcsname}

\DefineRemark{servercntonlyhandshakelast}{29}
\DefineRemark{serversinsecurecntwoaccess}{969}
\DefineRemark{serversinsecurepctwoaccess}{86}
\DefineRemark{serversinsecurecnt}{1030}
\DefineRemark{serversinsecurepct}{92}
\DefineRemark{serversaccessiblecnt}{493}
\DefineRemark{serversaccessiblepct}{24}
\DefineRemark{datasetfirstmeasurement}{2020-02-09}
\DefineRemark{datasetlastmeasurement}{2020-08-30}
\DefineRemark{datasetcntbeckhoffserverslast}{112}
\DefineRemark{datasetcntbachmannserverslast}{406}
\DefineRemark{datasetcntsiemensserverslast}{46}
\DefineRemark{datasetcntwagoserverslast}{78}
\DefineRemark{datasetcntallserverslast}{1953}
\DefineRemark{datasetminallservers}{1761}
\DefineRemark{datasetmaxallservers}{2069}
\DefineRemark{datasetcntdiscoveryserverslast}{839}
\DefineRemark{datasetpctdiscoveryserverslast}{42}
\DefineRemark{datasetcntnondiscoveryserverslast}{1114}
\DefineRemark{modescntallserverslast}{1114}
\DefineRemark{modescntNoneserverslast}{1035}
\DefineRemark{modescntSignserverslast}{588}
\DefineRemark{modescntSignAndEncryptserverslast}{843}
\DefineRemark{modespctNoneserverslast}{92}
\DefineRemark{modespctSignserverslast}{52}
\DefineRemark{modespctSignAndEncryptserverslast}{75}
\DefineRemark{modescntNoneLSMserverslast}{1035}
\DefineRemark{modescntNoneMSMserverslast}{270}
\DefineRemark{modescntSignLSMserverslast}{28}
\DefineRemark{modescntSignMSMserverslast}{1}
\DefineRemark{modescntSignAndEncryptLSMserverslast}{51}
\DefineRemark{modescntSignAndEncryptMSMserverslast}{843}
\DefineRemark{modespctNoneLSMserverslast}{100}
\DefineRemark{modespctNoneMSMserverslast}{26}
\DefineRemark{modespctSignLSMserverslast}{2.7}
\DefineRemark{modespctSignMSMserverslast}{0.1}
\DefineRemark{modespctSignAndEncryptLSMserverslast}{4.9}
\DefineRemark{modespctSignAndEncryptMSMserverslast}{81}
\DefineRemark{modescntsesLSMserverslast}{79}
\DefineRemark{modespctsesLSMserverslast}{7.1}
\DefineRemark{modescntsesMSMserverslast}{844}
\DefineRemark{modespctsesMSMserverslast}{75}
\DefineRemark{policiescntallserverslast}{1114}
\DefineRemark{policiescntNoneserverslast}{1035}
\DefineRemark{policiescntBasic128Rsa15serverslast}{715}
\DefineRemark{policiescntBasic256serverslast}{762}
\DefineRemark{policiescntAes128Sha256RsaOaepserverslast}{10}
\DefineRemark{policiescntBasic256Sha256serverslast}{564}
\DefineRemark{policiescntAes256Sha256RsaPssserverslast}{8}
\DefineRemark{policiespctNoneserverslast}{92}
\DefineRemark{policiespctBasic128Rsa15serverslast}{64}
\DefineRemark{policiespctBasic256serverslast}{68}
\DefineRemark{policiespctAes128Sha256RsaOaepserverslast}{0.9}
\DefineRemark{policiespctBasic256Sha256serverslast}{50}
\DefineRemark{policiespctAes256Sha256RsaPssserverslast}{0.7}
\DefineRemark{policiescntNoneLSPserverslast}{1035}
\DefineRemark{policiescntNoneMSPserverslast}{270}
\DefineRemark{policiescntBasic128Rsa15LSPserverslast}{13}
\DefineRemark{policiescntBasic128Rsa15MSPserverslast}{24}
\DefineRemark{policiescntBasic256LSPserverslast}{50}
\DefineRemark{policiescntBasic256MSPserverslast}{256}
\DefineRemark{policiescntAes128Sha256RsaOaepLSPserverslast}{0}
\DefineRemark{policiescntAes128Sha256RsaOaepMSPserverslast}{0}
\DefineRemark{policiescntBasic256Sha256LSPserverslast}{16}
\DefineRemark{policiescntBasic256Sha256MSPserverslast}{556}
\DefineRemark{policiescntAes256Sha256RsaPssLSPserverslast}{0}
\DefineRemark{policiescntAes256Sha256RsaPssMSPserverslast}{8}
\DefineRemark{policiespctNoneLSPserverslast}{100}
\DefineRemark{policiespctNoneMSPserverslast}{26}
\DefineRemark{policiespctBasic128Rsa15LSPserverslast}{1.3}
\DefineRemark{policiespctBasic128Rsa15MSPserverslast}{2.3}
\DefineRemark{policiespctBasic256LSPserverslast}{4.8}
\DefineRemark{policiespctBasic256MSPserverslast}{24}
\DefineRemark{policiespctAes128Sha256RsaOaepLSPserverslast}{0.0}
\DefineRemark{policiespctAes128Sha256RsaOaepMSPserverslast}{0.0}
\DefineRemark{policiespctBasic256Sha256LSPserverslast}{1.5}
\DefineRemark{policiespctBasic256Sha256MSPserverslast}{53}
\DefineRemark{policiespctAes256Sha256RsaPssLSPserverslast}{0.0}
\DefineRemark{policiespctAes256Sha256RsaPssMSPserverslast}{0.8}
\DefineRemark{policiescntofferhighserverslast}{564}
\DefineRemark{policiespctofferhighserverslast}{50}
\DefineRemark{policiescntenforcehighserverslast}{16}
\DefineRemark{policiespctenforcehighserverslast}{1.4}
\DefineRemark{policiescntofferdeprecatedserverslast}{786}
\DefineRemark{policiespctofferdeprecatedserverslast}{70}
\DefineRemark{policiescntenforcedeprecatedserverslast}{280}
\DefineRemark{policiespctenforcedeprecatedserverslast}{25}
\DefineRemark{policiescntenforcedeprecatedornoneserverslast}{550}
\DefineRemark{policiespctenforcedeprecatedornoneserverslast}{49}
\DefineRemark{takewaycnttheoreticallysufficientsecurity}{564}
\DefineRemark{policycntlevelstrongestmismatch}{19}
\DefineRemark{policycntlevelunusable}{18}
\DefineRemark{certsserverscntlast}{881}
\DefineRemark{certscntsigkeymd5WithRSAEncryption10240}{7}
\DefineRemark{certscntsigkeysha1WithRSAEncryption10240}{585}
\DefineRemark{certscntsigkeysha1WithRSAEncryption20480}{72}
\DefineRemark{certscntsigkeysha256WithRSAEncryption10240}{13}
\DefineRemark{certscntsigkeysha256WithRSAEncryption20480}{183}
\DefineRemark{certscntsigkeysha256WithRSAEncryption40960}{23}
\DefineRemark{certscntsigmd5WithRSAEncryption}{7}
\DefineRemark{certscntsigsha1WithRSAEncryption}{657}
\DefineRemark{certscntsigsha256WithRSAEncryption}{219}
\DefineRemark{certscntkeysize10240}{604}
\DefineRemark{certscntkeysize20480}{255}
\DefineRemark{certscntkeysize40960}{23}
\DefineRemark{certscntbasic256sha256all}{564}
\DefineRemark{certscntbasic256sha256nomatch}{409}
\DefineRemark{certscntaes256sha256rsapssall}{8}
\DefineRemark{certscntbasic128rsa15all}{701}
\DefineRemark{certscntbasic128rsa15nomatch}{82}
\DefineRemark{certscntbasic128rsa15toostrong}{75}
\DefineRemark{certscnthighpolicynomatch}{409}
\DefineRemark{certspcthighpolicynomatch}{72}
\DefineRemark{certsreusehostsmax}{385}
\DefineRemark{certsreuseasesmax}{24}
\DefineRemark{certsreusehostsmid}{9}
\DefineRemark{certsreuseasesmid}{8}
\DefineRemark{certsreusehostsmin}{6}
\DefineRemark{certsreuseasesmin}{5}
\DefineRemark{servercntsecurechanoklast}{1034}
\DefineRemark{servercntsecurechanokanonymouslast}{563}
\DefineRemark{serverpctsecurechanokanonymouslast}{50}
\DefineRemark{servercntcertoklast}{563}
\DefineRemark{servercntcertokaccessiblelast}{493}
\DefineRemark{serverpctcertokaccessiblelast}{44}
\DefineRemark{servercntcertokaccessfaillast}{70}
\DefineRemark{servercntaccessibleonlydefaultlast}{156}
\DefineRemark{servercntaccessibleproductivelast}{295}
\DefineRemark{serverpctaccessibleproductivelast}{26}
\DefineRemark{servercntaccessibletestlast}{42}
\DefineRemark{servercntaccessiblelast}{493}
\DefineRemark{serverpctaccessiblelast}{44}
\DefineRemark{servercntaccessibleproductiveanonlast}{116}
\DefineRemark{serverpctaccessibleproductiveanonlast}{10}
\DefineRemark{servercntaccessibleproductivecredlast}{0}
\DefineRemark{serverpctaccessibleproductivecredlast}{0.0}
\DefineRemark{servercntaccessibleproductivecertlast}{0}
\DefineRemark{serverpctaccessibleproductivecertlast}{0.0}
\DefineRemark{servercntaccessibleproductivetokenlast}{0}
\DefineRemark{serverpctaccessibleproductivetokenlast}{0.0}
\DefineRemark{servercntaccessibleproductiveanoncredlast}{168}
\DefineRemark{serverpctaccessibleproductiveanoncredlast}{15}
\DefineRemark{servercntaccessibleproductiveanoncertlast}{0}
\DefineRemark{serverpctaccessibleproductiveanoncertlast}{0.0}
\DefineRemark{servercntaccessibleproductiveanontokenlast}{0}
\DefineRemark{serverpctaccessibleproductiveanontokenlast}{0.0}
\DefineRemark{servercntaccessibleproductivecredcertlast}{0}
\DefineRemark{serverpctaccessibleproductivecredcertlast}{0.0}
\DefineRemark{servercntaccessibleproductivecredtokenlast}{0}
\DefineRemark{serverpctaccessibleproductivecredtokenlast}{0.0}
\DefineRemark{servercntaccessibleproductivecerttokenlast}{0}
\DefineRemark{serverpctaccessibleproductivecerttokenlast}{0.0}
\DefineRemark{servercntaccessibleproductiveanoncredcertlast}{11}
\DefineRemark{serverpctaccessibleproductiveanoncredcertlast}{1.0}
\DefineRemark{servercntaccessibleproductiveanoncredtokenlast}{0}
\DefineRemark{serverpctaccessibleproductiveanoncredtokenlast}{0.0}
\DefineRemark{servercntaccessibleproductiveanoncerttokenlast}{0}
\DefineRemark{serverpctaccessibleproductiveanoncerttokenlast}{0.0}
\DefineRemark{servercntaccessibleproductivecredcerttokenlast}{0}
\DefineRemark{serverpctaccessibleproductivecredcerttokenlast}{0.0}
\DefineRemark{servercntaccessibleproductiveanoncredcerttokenlast}{0}
\DefineRemark{serverpctaccessibleproductiveanoncredcerttokenlast}{0.0}
\DefineRemark{serverpctaccessibletestlast}{3.8}
\DefineRemark{servercntaccessibletestanonlast}{8}
\DefineRemark{serverpctaccessibletestanonlast}{0.7}
\DefineRemark{servercntaccessibletestcredlast}{0}
\DefineRemark{serverpctaccessibletestcredlast}{0.0}
\DefineRemark{servercntaccessibletestcertlast}{0}
\DefineRemark{serverpctaccessibletestcertlast}{0.0}
\DefineRemark{servercntaccessibletesttokenlast}{0}
\DefineRemark{serverpctaccessibletesttokenlast}{0.0}
\DefineRemark{servercntaccessibletestanoncredlast}{20}
\DefineRemark{serverpctaccessibletestanoncredlast}{1.8}
\DefineRemark{servercntaccessibletestanoncertlast}{0}
\DefineRemark{serverpctaccessibletestanoncertlast}{0.0}
\DefineRemark{servercntaccessibletestanontokenlast}{0}
\DefineRemark{serverpctaccessibletestanontokenlast}{0.0}
\DefineRemark{servercntaccessibletestcredcertlast}{0}
\DefineRemark{serverpctaccessibletestcredcertlast}{0.0}
\DefineRemark{servercntaccessibletestcredtokenlast}{0}
\DefineRemark{serverpctaccessibletestcredtokenlast}{0.0}
\DefineRemark{servercntaccessibletestcerttokenlast}{0}
\DefineRemark{serverpctaccessibletestcerttokenlast}{0.0}
\DefineRemark{servercntaccessibletestanoncredcertlast}{14}
\DefineRemark{serverpctaccessibletestanoncredcertlast}{1.3}
\DefineRemark{servercntaccessibletestanoncredtokenlast}{0}
\DefineRemark{serverpctaccessibletestanoncredtokenlast}{0.0}
\DefineRemark{servercntaccessibletestanoncerttokenlast}{0}
\DefineRemark{serverpctaccessibletestanoncerttokenlast}{0.0}
\DefineRemark{servercntaccessibletestcredcerttokenlast}{0}
\DefineRemark{serverpctaccessibletestcredcerttokenlast}{0.0}
\DefineRemark{servercntaccessibletestanoncredcerttokenlast}{0}
\DefineRemark{serverpctaccessibletestanoncredcerttokenlast}{0.0}
\DefineRemark{servercntaccessibleunclassifiedlast}{156}
\DefineRemark{serverpctaccessibleunclassifiedlast}{14}
\DefineRemark{servercntaccessibleunclassifiedanonlast}{5}
\DefineRemark{serverpctaccessibleunclassifiedanonlast}{0.4}
\DefineRemark{servercntaccessibleunclassifiedcredlast}{0}
\DefineRemark{serverpctaccessibleunclassifiedcredlast}{0.0}
\DefineRemark{servercntaccessibleunclassifiedcertlast}{0}
\DefineRemark{serverpctaccessibleunclassifiedcertlast}{0.0}
\DefineRemark{servercntaccessibleunclassifiedtokenlast}{0}
\DefineRemark{serverpctaccessibleunclassifiedtokenlast}{0.0}
\DefineRemark{servercntaccessibleunclassifiedanoncredlast}{134}
\DefineRemark{serverpctaccessibleunclassifiedanoncredlast}{12}
\DefineRemark{servercntaccessibleunclassifiedanoncertlast}{0}
\DefineRemark{serverpctaccessibleunclassifiedanoncertlast}{0.0}
\DefineRemark{servercntaccessibleunclassifiedanontokenlast}{0}
\DefineRemark{serverpctaccessibleunclassifiedanontokenlast}{0.0}
\DefineRemark{servercntaccessibleunclassifiedcredcertlast}{0}
\DefineRemark{serverpctaccessibleunclassifiedcredcertlast}{0.0}
\DefineRemark{servercntaccessibleunclassifiedcredtokenlast}{0}
\DefineRemark{serverpctaccessibleunclassifiedcredtokenlast}{0.0}
\DefineRemark{servercntaccessibleunclassifiedcerttokenlast}{0}
\DefineRemark{serverpctaccessibleunclassifiedcerttokenlast}{0.0}
\DefineRemark{servercntaccessibleunclassifiedanoncredcertlast}{17}
\DefineRemark{serverpctaccessibleunclassifiedanoncredcertlast}{1.5}
\DefineRemark{servercntaccessibleunclassifiedanoncredtokenlast}{0}
\DefineRemark{serverpctaccessibleunclassifiedanoncredtokenlast}{0.0}
\DefineRemark{servercntaccessibleunclassifiedanoncerttokenlast}{0}
\DefineRemark{serverpctaccessibleunclassifiedanoncerttokenlast}{0.0}
\DefineRemark{servercntaccessibleunclassifiedcredcerttokenlast}{0}
\DefineRemark{serverpctaccessibleunclassifiedcredcerttokenlast}{0.0}
\DefineRemark{servercntaccessibleunclassifiedanoncredcerttokenlast}{0}
\DefineRemark{serverpctaccessibleunclassifiedanoncredcerttokenlast}{0.0}
\DefineRemark{servercntunaccessiblelast}{621}
\DefineRemark{serverpctunaccessiblelast}{55}
\DefineRemark{servercntunaccessibleauthlast}{541}
\DefineRemark{serverpctunaccessibleauthlast}{48}
\DefineRemark{servercntunaccessibleauthanonlast}{9}
\DefineRemark{serverpctunaccessibleauthanonlast}{0.8}
\DefineRemark{servercntunaccessibleauthcredlast}{464}
\DefineRemark{serverpctunaccessibleauthcredlast}{41}
\DefineRemark{servercntunaccessibleauthcertlast}{0}
\DefineRemark{serverpctunaccessibleauthcertlast}{0.0}
\DefineRemark{servercntunaccessibleauthtokenlast}{0}
\DefineRemark{serverpctunaccessibleauthtokenlast}{0.0}
\DefineRemark{servercntunaccessibleauthanoncredlast}{38}
\DefineRemark{serverpctunaccessibleauthanoncredlast}{3.4}
\DefineRemark{servercntunaccessibleauthanoncertlast}{0}
\DefineRemark{serverpctunaccessibleauthanoncertlast}{0.0}
\DefineRemark{servercntunaccessibleauthanontokenlast}{0}
\DefineRemark{serverpctunaccessibleauthanontokenlast}{0.0}
\DefineRemark{servercntunaccessibleauthcredcertlast}{4}
\DefineRemark{serverpctunaccessibleauthcredcertlast}{0.4}
\DefineRemark{servercntunaccessibleauthcredtokenlast}{0}
\DefineRemark{serverpctunaccessibleauthcredtokenlast}{0.0}
\DefineRemark{servercntunaccessibleauthcerttokenlast}{0}
\DefineRemark{serverpctunaccessibleauthcerttokenlast}{0.0}
\DefineRemark{servercntunaccessibleauthanoncredcertlast}{17}
\DefineRemark{serverpctunaccessibleauthanoncredcertlast}{1.5}
\DefineRemark{servercntunaccessibleauthanoncredtokenlast}{0}
\DefineRemark{serverpctunaccessibleauthanoncredtokenlast}{0.0}
\DefineRemark{servercntunaccessibleauthanoncerttokenlast}{0}
\DefineRemark{serverpctunaccessibleauthanoncerttokenlast}{0.0}
\DefineRemark{servercntunaccessibleauthcredcerttokenlast}{0}
\DefineRemark{serverpctunaccessibleauthcredcerttokenlast}{0.0}
\DefineRemark{servercntunaccessibleauthanoncredcerttokenlast}{6}
\DefineRemark{serverpctunaccessibleauthanoncredcerttokenlast}{0.5}
\DefineRemark{servercntunaccessiblesecchanlast}{80}
\DefineRemark{serverpctunaccessiblesecchanlast}{7.2}
\DefineRemark{servercntunaccessiblesecchananonlast}{1}
\DefineRemark{serverpctunaccessiblesecchananonlast}{0.1}
\DefineRemark{servercntunaccessiblesecchancredlast}{21}
\DefineRemark{serverpctunaccessiblesecchancredlast}{1.9}
\DefineRemark{servercntunaccessiblesecchancertlast}{0}
\DefineRemark{serverpctunaccessiblesecchancertlast}{0.0}
\DefineRemark{servercntunaccessiblesecchantokenlast}{0}
\DefineRemark{serverpctunaccessiblesecchantokenlast}{0.0}
\DefineRemark{servercntunaccessiblesecchananoncredlast}{5}
\DefineRemark{serverpctunaccessiblesecchananoncredlast}{0.4}
\DefineRemark{servercntunaccessiblesecchananoncertlast}{0}
\DefineRemark{serverpctunaccessiblesecchananoncertlast}{0.0}
\DefineRemark{servercntunaccessiblesecchananontokenlast}{0}
\DefineRemark{serverpctunaccessiblesecchananontokenlast}{0.0}
\DefineRemark{servercntunaccessiblesecchancredcertlast}{7}
\DefineRemark{serverpctunaccessiblesecchancredcertlast}{0.6}
\DefineRemark{servercntunaccessiblesecchancredtokenlast}{0}
\DefineRemark{serverpctunaccessiblesecchancredtokenlast}{0.0}
\DefineRemark{servercntunaccessiblesecchancerttokenlast}{0}
\DefineRemark{serverpctunaccessiblesecchancerttokenlast}{0.0}
\DefineRemark{servercntunaccessiblesecchananoncredcertlast}{3}
\DefineRemark{serverpctunaccessiblesecchananoncredcertlast}{0.3}
\DefineRemark{servercntunaccessiblesecchananoncredtokenlast}{43}
\DefineRemark{serverpctunaccessiblesecchananoncredtokenlast}{3.9}
\DefineRemark{servercntunaccessiblesecchananoncerttokenlast}{0}
\DefineRemark{serverpctunaccessiblesecchananoncerttokenlast}{0.0}
\DefineRemark{servercntunaccessiblesecchancredcerttokenlast}{0}
\DefineRemark{serverpctunaccessiblesecchancredcerttokenlast}{0.0}
\DefineRemark{servercntunaccessiblesecchananoncredcerttokenlast}{0}
\DefineRemark{serverpctunaccessiblesecchananoncredcerttokenlast}{0.0}
\DefineRemark{servercntauthanonlast}{139}
\DefineRemark{serverpctauthanonlast}{12}
\DefineRemark{servercntauthcredlast}{485}
\DefineRemark{serverpctauthcredlast}{43}
\DefineRemark{servercntauthcertlast}{0}
\DefineRemark{serverpctauthcertlast}{0.0}
\DefineRemark{servercntauthtokenlast}{0}
\DefineRemark{serverpctauthtokenlast}{0.0}
\DefineRemark{servercntauthanoncredlast}{365}
\DefineRemark{serverpctauthanoncredlast}{32}
\DefineRemark{servercntauthanoncertlast}{0}
\DefineRemark{serverpctauthanoncertlast}{0.0}
\DefineRemark{servercntauthanontokenlast}{0}
\DefineRemark{serverpctauthanontokenlast}{0.0}
\DefineRemark{servercntauthcredcertlast}{11}
\DefineRemark{serverpctauthcredcertlast}{1.0}
\DefineRemark{servercntauthcredtokenlast}{0}
\DefineRemark{serverpctauthcredtokenlast}{0.0}
\DefineRemark{servercntauthcerttokenlast}{0}
\DefineRemark{serverpctauthcerttokenlast}{0.0}
\DefineRemark{servercntauthanoncredcertlast}{62}
\DefineRemark{serverpctauthanoncredcertlast}{5.6}
\DefineRemark{servercntauthanoncredtokenlast}{43}
\DefineRemark{serverpctauthanoncredtokenlast}{3.9}
\DefineRemark{servercntauthanoncerttokenlast}{0}
\DefineRemark{serverpctauthanoncerttokenlast}{0.0}
\DefineRemark{servercntauthcredcerttokenlast}{0}
\DefineRemark{serverpctauthcredcerttokenlast}{0.0}
\DefineRemark{servercntauthanoncredcerttokenlast}{6}
\DefineRemark{serverpctauthanoncredcerttokenlast}{0.5}
\DefineRemark{accesspctrx}{90}
\DefineRemark{accesspctry}{97}
\DefineRemark{accesspctwx}{33}
\DefineRemark{accesspctwy}{10}
\DefineRemark{accesspctex}{61}
\DefineRemark{accesspctey}{86}
\DefineRemark{nodepctaccessanonymousreadlast}{82}
\DefineRemark{nodepctaccessanonymouswritelast}{46}
\DefineRemark{nodecntvarproductive}{239824}
\DefineRemark{nodecntvarproductiveanonread}{235459}
\DefineRemark{nodepctvarproductiveanonread}{98}
\DefineRemark{nodepctvarproductiveanonreaddefaultns}{94}
\DefineRemark{nodepctvarproductiveanonreadcustomns}{98}
\DefineRemark{nodepctvarproductiveanonwrite}{59}
\DefineRemark{nodepctvarproductiveanonwritecustomns}{64}
\DefineRemark{productiveserverscntfuncexecutable}{161}
\DefineRemark{productiveserverspctfuncexecutable}{54}
\DefineRemark{nodepctanonymousmethodexecutable}{86}
\DefineRemark{datasetpctopcuahandshake}{0.5}
\DefineRemark{datasetcntpotentialduplicatelast}{17}
\DefineRemark{datasetpctpotentialduplicatelast}{0.9}
\DefineRemark{certcntreusebachmannlast}{385}
\DefineRemark{certcntreusebachmannlastuniqueases}{24}
\DefineRemark{certcntreusebachmannlastdate}{2020-08-30}
\DefineRemark{certcntreusebachmannmax}{387}
\DefineRemark{certcntreusebachmannmaxuniqueases}{24}
\DefineRemark{certcntreusebachmannmaxdate}{2020-08-09}
\DefineRemark{certcntreusebachmannmin}{263}
\DefineRemark{certcntreusebachmannminuniqueases}{17}
\DefineRemark{certcntreusebachmannmindate}{2020-02-09}
\DefineRemark{certcntreuseatvise1last}{9}
\DefineRemark{certcntreuseatvise1lastuniqueases}{8}
\DefineRemark{certcntreuseatvise1lastdate}{2020-08-30}
\DefineRemark{certcntreuseatvise1max}{13}
\DefineRemark{certcntreuseatvise1maxuniqueases}{10}
\DefineRemark{certcntreuseatvise1maxdate}{2020-04-05}
\DefineRemark{certcntreuseatvise1min}{7}
\DefineRemark{certcntreuseatvise1minuniqueases}{6}
\DefineRemark{certcntreuseatvise1mindate}{2020-08-02}
\DefineRemark{certcntreuseatvise2last}{6}
\DefineRemark{certcntreuseatvise2lastuniqueases}{5}
\DefineRemark{certcntreuseatvise2lastdate}{2020-08-30}
\DefineRemark{certcntreuseatvise2max}{9}
\DefineRemark{certcntreuseatvise2maxuniqueases}{6}
\DefineRemark{certcntreuseatvise2maxdate}{2020-03-01}
\DefineRemark{certcntreuseatvise2min}{4}
\DefineRemark{certcntreuseatvise2minuniqueases}{4}
\DefineRemark{certcntreuseatvise2mindate}{2020-05-10}
\DefineRemark{servercntbachmannweeklyaddavg}{4}
\DefineRemark{servercntbachmannweeklyaddlast}{3}
\DefineRemark{servercntbachmannaddsincemay5}{24}
\DefineRemark{certspctselfsignedlast}{99}
\DefineRemark{takeawaycntnonelast}{270}
\DefineRemark{takeawaypctnonelast}{24}
\DefineRemark{takeawaycntnonedeprecatedlast}{550}
\DefineRemark{takeawaypctnonedeprecatedlast}{49}
\DefineRemark{takeawaycntnonedeprecatedlastadd}{280}
\DefineRemark{takeawaypctnonedeprecatedlastadd}{25}
\DefineRemark{takeawaycntnonedeprecatedtooweaklast}{947}
\DefineRemark{takeawaypctnonedeprecatedtooweaklast}{85}
\DefineRemark{takeawaycntnonedeprecatedtooweaklastadd}{397}
\DefineRemark{takeawaypctnonedeprecatedtooweaklastadd}{35}
\DefineRemark{takeawaycntnonedeprecatedtooweakreusedlast}{952}
\DefineRemark{takeawaypctnonedeprecatedtooweakreusedlast}{85}
\DefineRemark{takeawaycntnonedeprecatedtooweakreusedlastadd}{5}
\DefineRemark{takeawaypctnonedeprecatedtooweakreusedlastadd}{0.4}
\DefineRemark{takeawaycntnonedeprecatedtooweakreusedaccessiblelast}{1023}
\DefineRemark{takeawaypctnonedeprecatedtooweakreusedaccessiblelast}{91}
\DefineRemark{takeawaycntnonedeprecatedtooweakreusedaccessiblelastadd}{71}
\DefineRemark{takeawaypctnonedeprecatedtooweakreusedaccessiblelastadd}{6.4}
\DefineRemark{takeawaycntdeprecatedlast}{280}
\DefineRemark{takeawaypctdeprecatedlast}{25}
\DefineRemark{takeawaycntdeprecatedtooweaklast}{677}
\DefineRemark{takeawaypctdeprecatedtooweaklast}{60}
\DefineRemark{takeawaycntdeprecatedtooweaklastadd}{397}
\DefineRemark{takeawaypctdeprecatedtooweaklastadd}{35}
\DefineRemark{takeawaycntdeprecatedtooweakreusedlast}{682}
\DefineRemark{takeawaypctdeprecatedtooweakreusedlast}{61}
\DefineRemark{takeawaycntdeprecatedtooweakreusedlastadd}{5}
\DefineRemark{takeawaypctdeprecatedtooweakreusedlastadd}{0.4}
\DefineRemark{takeawaycntdeprecatedtooweakreusedaccessiblelast}{753}
\DefineRemark{takeawaypctdeprecatedtooweakreusedaccessiblelast}{67}
\DefineRemark{takeawaycntdeprecatedtooweakreusedaccessiblelastadd}{71}
\DefineRemark{takeawaypctdeprecatedtooweakreusedaccessiblelastadd}{6.4}
\DefineRemark{takeawaycntnoneaccessiblelast}{341}
\DefineRemark{takeawaypctnoneaccessiblelast}{30}
\DefineRemark{takeawaycntnoneaccessiblelastadd}{71}
\DefineRemark{takeawaypctnoneaccessiblelastadd}{6.4}
\DefineRemark{takeawaycnttheoreticallysecurelast}{844}
\DefineRemark{takeawaycntpropersecuritybutaccessiblelast}{71}
\DefineRemark{serverspctdeficitavg}{92}
\DefineRemark{serverspctdeficitmin}{91}
\DefineRemark{serverspctdeficitmax}{94}
\DefineRemark{serverspctdeficitstd}{0.8}
\DefineRemark{certscntall}{4296}
\DefineRemark{certscntsha1issuedsince2019}{1923}
\DefineRemark{certspctsha1issuedsince2019}{44}
\DefineRemark{certscntsha1issuedbefore2019}{1711}
\DefineRemark{certspctsha1issuedbefore2019}{39}
\DefineRemark{certscntsha1issuedafter2017}{2174}
\DefineRemark{certspctsha1issuedafter2017}{50}
\DefineRemark{certspctsha1issuedsince2019frac2017}{88}
\DefineRemark{certscntsha256issuedsince2019}{1340}
\DefineRemark{certspctsha256issuedsince2019}{31}
\DefineRemark{certscntsha256issuedbefore2019}{202}
\DefineRemark{certspctsha256issuedbefore2019}{4.7}
\DefineRemark{certscntkl1024issuedafter2017}{2321}
\DefineRemark{certspctkl1024issuedafter2017}{54}
\DefineRemark{certscntsha1issued2019last}{203}
\DefineRemark{certscntreusedlast}{9}
\DefineRemark{measurementtimeavglast}{110}
\DefineRemark{measurementtimemsminlast}{31}
\DefineRemark{measurementtimemaxlast}{5393}
\DefineRemark{measurementtimestdlast}{461}
\DefineRemark{measurementtraffickbavglast}{1015}
\DefineRemark{measurementtrafficbminlast}{218}
\DefineRemark{measurementtrafficmbmaxlast}{104}
\DefineRemark{measurementtrafficmbstdlast}{6}
\DefineRemark{measurementreadkbavglast}{352}
\DefineRemark{measurementreadbminlast}{28}
\DefineRemark{measurementreadmbmaxlast}{50}
\DefineRemark{measurementreadmbstdlast}{2}
\DefineRemark{measurementwrittenkbavglast}{662}
\DefineRemark{measurementwrittenbminlast}{190}
\DefineRemark{measurementwrittenmbmaxlast}{67}
\DefineRemark{measurementwrittenmbstdlast}{4}

\begin{document}

\title[An Internet-Wide Study on Insecure OPC UA Deployments]{Easing the Conscience with OPC UA:\\An Internet-Wide Study on Insecure Deployments}

\ifmanualheader
  \author{Markus Dahlmanns\(^*\), Johannes Lohmöller\(^*\), Ina Berenice Fink\(^*\),\\ Jan Pennekamp\(^*\), Klaus Wehrle\(^*\), Martin Henze\(^\ddagger\)}
  \def\cleanauthors{Markus Dahlmanns, Johannes Lohmöller, Ina Berenice Fink, Jan Pennekamp, Klaus Wehrle, Martin Henze}
  \affiliation{
  \(^*\)\textit{Communication and Distributed Systems}, RWTH Aachen University, Aachen, Germany \\
  \(^\ddagger\)\textit{Cyber Analysis \& Defense}, Fraunhofer FKIE, Wachtberg, Germany\\
  \{dahlmanns, lohmoeller, fink, pennekamp, wehrle\}@comsys.rwth-aachen.de \(\cdot\)
  martin.henze@fkie.fraunhofer.de
  }
\fi

\renewcommand{\shortauthors}{Dahlmanns et al.}

\begin{abstract}
    Due to increasing digitalization, formerly isolated industrial networks, e.g., for factory and process automation, move closer and closer to the Internet, mandating secure communication.
    However, securely setting up OPC~UA, the prime candidate for secure industrial communication, is challenging due to a large variety of insecure options.
    To study whether Internet-facing OPC~UA appliances are configured securely, we actively scan the IPv4 address space for publicly reachable OPC~UA systems and assess the security of their configurations.
    We observe problematic security configurations such as missing access control~(on \Remark{serversaccessiblepct}\% of hosts), disabled security functionality~(\Remark{takeawaypctnonelast}\%), or use of deprecated cryptographic primitives~(\Remark{takeawaypctdeprecatedlast}\%) on in total \Remark{serversinsecurepct}\% of the reachable deployments.
    Furthermore, we discover several hundred devices in multiple autonomous systems sharing the same security certificate, opening the door for impersonation attacks.
    Overall, in this paper, we highlight commonly found security misconfigurations and underline the importance of appropriate configuration for security-featuring protocols.
\end{abstract}



\maketitle

\section{Introduction}
Industrial networks, e.g., used for factory and process automation, traditionally were designed as isolated networks with no connections to, e.g., office networks or the Internet~\cite{mirian-icsmes-2016,cheminod-industrialsecurityissues-2013}.
Consequently, industrial protocols, such as Modbus or ProfiNet, do not implement any security functionality.
However, with an increasing interconnection of industrial networks, serious security threats arise as evidenced by incidents such as NotPetya or manipulation attacks on several industrial devices~\cite{hemsley-icsattacks-2018}.
These threats, coupled with an increase in industrial communication (e.g., driven by Industry~4.0), highlight the need for secure industrial protocols.

\emph{OPC~UA}, a comparatively new industrial protocol, released in 2008, was designed from scratch with security in mind~\cite{bsi-opcua-analysis} and is attested secure~(e.g., by the German Federal Office for Information Security~\cite{bsi-opcua-analysis}).
However, OPC~UA requires an active configuration of numerous security settings, where incautious decisions lead to weakly or even unsecured systems.
In industrial deployments, such configurations not only allow for well-known attacks, e.g., eavesdropping and theft of confidential data, but also facilitate to control production lines, cause physical damage, and harm humans~\cite{Henze2020The}.
Configuration recommendations~\cite{opcua-recommendation}, e.g., on the use of ciphers, attempt to confine the spread of insecure deployments.

However, until now, it is unclear whether system operators adhere to such security recommendations and therefore prevent unauthorized access to modern industrial deployments.
In other domains, active Internet-wide scanning has proven to be a valuable and accepted method to perform this task~\cite{gasser-ssh-2014, holz-tlscomm-2016,samarasinghe-anotherlooktls-2019,springall-tlsshortcuts-2016}.
Likewise, different works identify the risks of Internet-connected industrial devices using legacy protocols without security functionality~\cite{mirian-icsmes-2016,feng-characterizing-2016,bodenheim-shodaneval-2014}.
This motivates us to combine these two streams of research to analyze the security configurations of industrial deployments.

In this paper, we study whether Internet-connected OPC~UA deployments and their configurations capitalize on the strong level of security theoretically provided by the underlying protocol design~\cite{bsi-opcua-analysis}.
To this end, we actively scan the complete IPv4 address space for publicly reachable OPC~UA systems and subsequently assess the security configurations of found deployments.

\noindent\textbf{Contributions.}
Our main contributions are as follows.

\vspace{-0.55em}
\begin{itemize}[noitemsep,topsep=5pt,leftmargin=9pt]
    \item We perform weekly measurements of the complete IPv4 address space over seven months to detect OPC~UA devices, which we can attribute to well-known industrial manufacturers and sectors, e.g., building automation and power systems.
    \item We assess the security configurations of Internet-facing OPC~UA devices following official security guidelines and recommendations.
    Our results show that \Remark{serversinsecurepct}\% of OPC~UA systems are configured deficiently, e.g., due to missing access control, disabled security functionality, use of deprecated cryptographic primitives, or certificate reuse.
    \item We release our anonymized dataset~\cite{dahlmanns-opcuadataset-2020} and our OPC~UA extensions of \texttt{zgrab2}~\cite{opcua-github-code} to allow for reproducibility of our results.%
\end{itemize}

\section{A Primer on OPC UA and Security}
\label{sec:background}

Besides many functional improvements over prior industrial protocols, e.g., cross-vendor communication and platform independence, OPC~UA is the first widely-deployed industrial protocol with built-in and attested security~\cite{bsi-opcua-analysis}, allowing for secure remote access.

To enable platform-independent communication between industrial devices of different manufacturers, OPC~UA servers represent device functions, sensor values, and other variables as well as their relationships as a set of nodes in an address space, where namespaces provide semantic information about nodes~\cite{mahnke2009opc}.
From this address space, clients can dynamically request the execution of functions or data of variables.
While OPC~UA offers a variety of different communication paradigms and interfaces, e.g., publish/subscribe or HTTP(S), we focus on the binary interface (standard port 4840 via TCP) as it is a mandatory feature of all OPC~UA devices~\cite{opcua-profiles-2017}.
It implements security-specific mechanisms, such as authentication, access control, as well as integrity protection and confidentiality~\cite{opcua-security-2018}.

\begin{figure}[t]
    \centering
    \includegraphics[width=\linewidth]{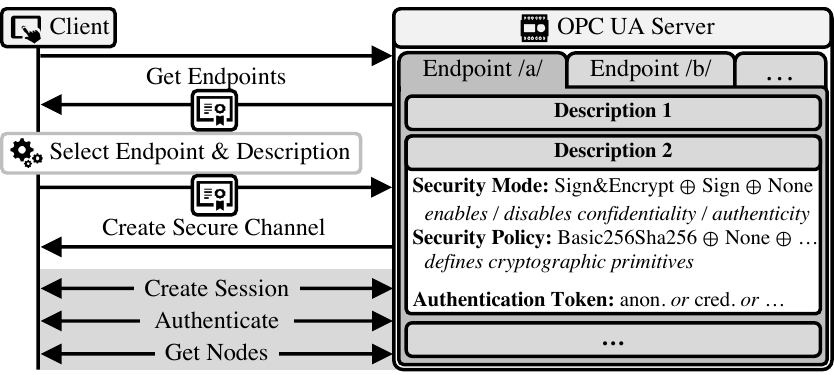}
    \vspace{-2em}
    \caption{%
        OPC~UA server configuration and communication steps with clients selecting one of multiple endpoints with different descriptions a server offers~(according to~\cite{opcua-security-2018,opcua-services-2017,opcua-profiles-2017}).
    }
    \vspace{-1.5em}
    \label{fig:opcua_background}
\end{figure}

Figure~\ref{fig:opcua_background} illustrates the establishment of a (secure) connection in OPC~UA, specifically focusing on information such as security primitives that an OPC~UA server provides to clients.
First, as OPC~UA servers can provide data via different endpoints, clients request a list of endpoints with a description of their security configurations.
Apart from this list, the response includes a certificate that authenticates the server.
After choosing one of the available endpoints, the client starts to establish a \textit{secure channel} using the channel parameters in the endpoint description.

Thereby, the \textit{security mode} in the endpoint description enables or disables confidentiality and/or authentication for communication.
Notably, establishing a secure channel already allows the client to authenticate to the server via a certificate, optionally realizing the first level of access control.
Table~\ref{tab:policies} lists the \textit{security policies} defining the cryptographic primitives for the secure channel establishment in the endpoint description, e.g., hash function and key length.

After establishing a secure channel, clients create a \textit{session} for subsequent data transmissions.
To access a server's address space, clients need to authenticate (using user credentials, a certificate, or an authentication token) unless anonymous access is enabled.
Depending on the authenticated user, OPC~UA servers can enforce different access control rules at the granularity of single nodes.

\begin{table}[t]
    \footnotesize
    \begin{tabularx}{\linewidth}{X|l|l|l|c}
    \textbf{Policy}                & \textbf{Sig. Hash} & \textbf{Cert. Hash}                           & \textbf{Key Len.} {[}\si{\bit}{]}     & \textbf{A} \\ \hline \hline
    \rowcolor[HTML]{B2BEB5} 
    None                  & ---       & ---          & ---              & N     \\
    \rowcolor[HTML]{B2BEB5} 
    Basic128Rsa15         & SHA1      & SHA1         & {[}1024; 2048{]} & D1    \\
    \rowcolor[HTML]{B2BEB5} 
    Basic256              & SHA1      & SHA1, SHA256 & {[}1024; 2048{]} & D2    \\
    Aes128\_Sha256\_RsaOaep & SHA256    & SHA256                               & {[}2048; 4096{]} & S1    \\
    Basic256Sha256        & SHA256    & SHA256                               & {[}2048; 4096{]} & S2    \\
    Aes256\_Sha256\_RsaPss  & SHA256    & SHA256                               & {[}2048; 4096{]} & S3   
    \end{tabularx}
    
    \hfill A:\ abbreviations for policies in the remainder of this paper
    \vspace{0.4em}
    \caption{OPC~UA security policies define used ciphers and key lengths (insecure and deprecated polices marked gray).}
    \vspace{-3em}
    \label{tab:policies}
\end{table}

Although OPC~UA's protocol design is secure~\cite{bsi-opcua-analysis}, its various configuration options can significantly impact security.
Official recommendations attempt to improve this situation~\cite{opcua-recommendation}:
First, \mbox{communication} security should never be disabled, i.e., signed and encrypted communication should be used whenever possible.
Likewise, anonymous authentication should be forbidden.
Finally, only three of the six available security policies should be used, as one provides no security and two have been deprecated due to the use of SHA-1~(cf. Table~\ref{tab:policies}).
Consequently, it is important to verify that OPC~UA deployments follow these recommendations.

\section{Related Work}

The benefits of a global view on the security configuration of OPC~UA deployments are emphasized by research on the security of Internet-facing industrial appliances as well as Internet-wide security analyses for Web protocols.

\textbf{Security of Industrial Deployments:}
Different works identify security issues of industrial deployments~\cite{hemsley-icsattacks-2018,quarta2017experimental,stellios2018survey,humayed-survey-2017,cheminod-industrialsecurityissues-2013, sadeghi-secprivchallengesiiot-2015}.
While actual security incidents are seldom~\cite{miller2012survey}, already a single incident can be catastrophic~\cite{nawrocki-passics-20}.
Remarkably, Mirian et al.~\cite{mirian-icsmes-2016} still found ten-thousands of industrial devices connected to the Internet via legacy and insecure protocols.
These devices can be classified, e.g., as programmable logic controllers~\cite{feng-characterizing-2016}.
Other research found robots controllable via the Internet~\cite{demarinis-rosscanning-2019}.

Internet scan projects, e.g., Censys~\cite{durumeric2015search} or Shodan~\cite{shodan}, offer meta-information about all Internet-connected devices, including industrial deployments~\cite{leverett-shodanclassification-2011, hansson-analysisshodan-2018}, detecting new industrial devices within one month~\cite{bodenheim-shodaneval-2014}.
Several studies analyzed Shodan's data on industrial devices to assess their security and found a vast amount of devices affected by known software vulnerabilities, e.g., in the Netherlands~\cite{ceron2019online}, in Finland~\cite{kiravuo-vulnerabilitiesfinland-2015}, and worldwide~\cite{genge-shovat-2016}.
These works only consider industrial devices using legacy industrial protocols.

Devices using these legacy and insecure industrial protocols are often subject to scanning activities~\cite{fachkha-cpsprobing-2017,husak-cpscyberawareness-2018,cabana-detectics-attacks-2019}.
While standard scanning tools, e.g., \texttt{zmap}~\cite{durumeric-zmap-2013}, not necessarily influence normal operations of industrial devices~\cite{coffey-scanninganalysis-2018}, malicious activities can compromise such unprotected devices.

Nawrocki et al.\ observe communication over legacy industrial protocols on IXP level and show that 96\% of messages originate from industrial devices~\cite{nawrocki-passics-20}, emphasizing the need for secure industrial communication.
Assessment guidelines and tools assist operators in correctly configuring secure industrial protocols such as OPC~UA~\cite{roepert-opcuaassessment-2020,holik_pentest_2014,opcua-recommendation}.

However, until now, it is an open question whether OPC~UA deployments actually capitalize on increased security functionality compared to legacy industrial protocols that provide no security.

\textbf{Internet Security Measurements:}
Active and passive measurements have proven useful for insights on the deployment and (mis-)configuration of security protocols.

Different works examine TLS deployments, i.e., the TLS and certificate configuration of Internet-facing embedded devices~\cite{samarasinghe-anotherlooktls-2019}, the spread of flaws in key generation~\cite{heninger-mining-2012} or TLS implementations~\cite{springall-tlsshortcuts-2016}, and the shift to newer versions and features~\cite{amann-tlsfeatures-2017}.
Further, related work analyzed security certificates regarding configuration~\cite{holz-509pki-2011}, validity~\cite{chung-certificates-2016}, wrong issuance~\cite{kumar-certificates-2018}, and certificate transparency logs~\cite{gasser-ctpractices-2018}.
Besides HTTPS deployments, also communication services~\cite{holz-tlscomm-2016} and SSH were examined~\cite{gasser-ssh-2014}.

In addition, Internet measurements have been utilized to find insecurely configured embedded devices~\cite{cui-insecureembedded-2010}, detect compromised IoT devices~\cite{shaikh-correlating-2018, neshenko-survey-2019, mangino-insecurity-2020}, and study cloud usage as well as communication security of IoT devices~\cite{ren-informationexposure-2019}.
Motivated by these observations, we set out to study whether modern Internet-facing industrial appliances that use the OPC~UA protocol capitalize on the promised increases in security.

\section{OPC UA Scanning \& Dataset}
\label{sec:scanning_dataset}

Our methodology to analyze the security configuration of Internet-facing OPC~UA deployments relies on weekly scans of the complete IPv4 address space on the default OPC~UA binary protocol port~(TCP, 4840), as implementing the binary protocol is mandatory for all OPC~UA instances~\cite{opcua-profiles-2017}.
By performing weekly scans, we can observe situation changes, e.g., software updates and certificate renewals.
During both, design and execution, we follow principles of ethical research and established best practices for Internet-wide active measurements~(cf.\ Appendix~\ref{sec:ethics}).
Whenever possible, we inform operators of insecure systems to prevent future harm.

\textbf{Scanner:}
We rely on \texttt{zmap} to detect Internet-facing systems with an open TCP port~4840 and \texttt{zgrab2}, which we extended with OPC~UA functionality based on \texttt{gopcua}, to connect to the found servers~(\Remark{datasetpctopcuahandshake}\textperthousand{} of hosts with an open TCP port~4840 actually run OPC~UA).
Subsequently, we retrieve information on the provided endpoints, their security configuration, i.e., available security modes, security policies, and authentication tokens, as well as establish a secure channel.
Whenever a server offers the security policy \texttt{Sign} or \texttt{SignAndEncrypt}, we send a self-signed certificate during the secure channel handshake.
Furthermore, we also connect to other host~/~port combinations listed as endpoints on scanned OPC~UA servers~(as of 2020-05-04).
For each server with anonymous access enabled, we traverse through the offered address space to retrieve all nodes and their access rights.

We use the collected information to
(i)~validate compliance of security configurations with security recommendations~\cite{opcua-recommendation},
(ii)~compare the conformance of advertised security configurations with used cryptographic ciphers, and
(iii)~discover devices neglecting security best practices~\cite{nist-sha1, nist-privatekey-2019}.

\begin{figure}[t]
  \centering
  \includegraphics[width=\linewidth]{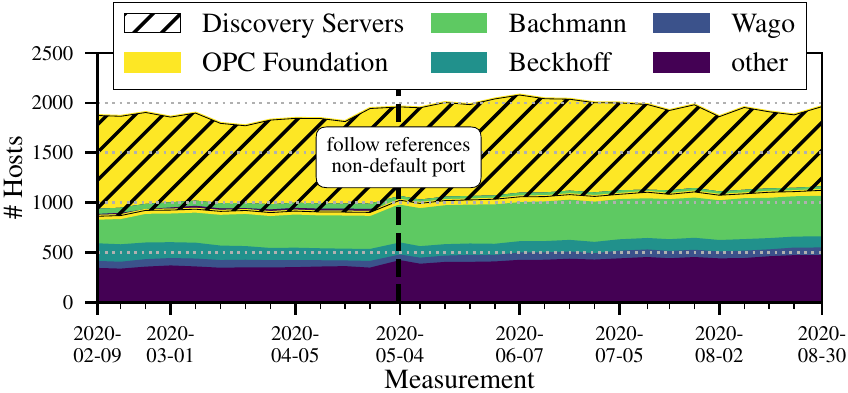}
  \vspace{-2em}
  \caption{OPC~UA servers found by our measurements can often be assigned to well-known manufacturers.}
  \vspace{-1.5em}
  \label{fig:dataset_description}
\end{figure}

\textbf{Dataset Overview:}
In Figure~\ref{fig:dataset_description}, we detail the number of publicly reachable OPC~UA servers over time~(seven months between \Remark{datasetfirstmeasurement} and \Remark{datasetlastmeasurement}), grouped by manufacturers of the software resp.\ industrial device (we manually clustered the values of the \texttt{ApplicationURI} field provided by servers).  %
We discovered between \Remark{datasetminallservers} and \Remark{datasetmaxallservers} deployments in the IPv4 address space. %

The found OPC~UA servers can be broadly classified into two categories.
First, \emph{discovery servers} (hatched in Figure~\ref{fig:dataset_description}, \Remark{datasetpctdiscoveryserverslast}\% of hosts in our latest measurement) only announcing OPC~UA endpoints running on other host~/~port combinations.
These systems mainly rely on the OPC~UA reference implementation~\cite{opcua-reference-implementation} and their \mbox{number} varies slightly from measurement to measurement. %
The second class comprises \emph{full OPC~UA servers}~(``servers'' from here on), mostly on industrial devices, which we could attribute to well-known industrial device manufacturers, most prominently Bachmann~(\Remark{datasetcntbachmannserverslast} devices in our last measurement), Beckhoff~(\Remark{datasetcntbeckhoffserverslast}), and Wago~(\Remark{datasetcntwagoserverslast}).
Their number marginally increased during the time of our measurements.
Since the security configuration of OPC~UA servers is only relevant for data transmission and not for the unprotected retrieval of endpoints, %
we focus on non-discovery servers below~(\Remark{datasetcntnondiscoveryserverslast}).

To enable reproducibility of our results, we make our anonymized dataset~\cite{dahlmanns-opcuadataset-2020} and extensions of \texttt{zgrab2}~\cite{opcua-github-code} publicly available.

\section{Security of OPC UA Deployments}

While, in principle a secure protocol, OPC~UA requires complex and correct configuration to achieve a secure deployment~\cite{bsi-opcua-analysis}.
Configuration recommendations~\cite{opcua-recommendation} as well as general security advice for industrial control systems~\cite{nist-icsrecommendations-2015} and generic security guidelines, e.g., on the use of certificates~\cite{opcua-recommendation, nist-privatekey-2019} and hash functions~\cite{nist-sha1,bsi-sha1}, aid this task.
However, it is unclear whether operators follow these recommendations to secure their deployments.
To overcome this gap of knowledge, we analyze and assess all reachable OPC~UA servers' security configurations w.r.t.\ communication security, authorization, and access control.
Unless stated otherwise, we rely on our latest measurement (\Remark{datasetlastmeasurement}) for our analysis.
In Appendix~\ref{sec:closerlook} we elaborate on the distribution of security configurations over device manufacturers and autonomous systems.

\subsection{Advertised Security Properties}

Using their endpoint descriptions, OPC~UA servers advertise different security modes and security policies and thus define foundational security settings and cryptographic primitives. %

\textbf{Message Security Mode:}
The client-chosen message security mode determines whether the communication is authenticated and confidential (mode~\texttt{SignAndEncrypt}), authenticated~(\texttt{Sign}), or insecure~(\texttt{None})~(cf.\ Section~\ref{sec:background}).
To enable clients to benefit from secure connections, it is fundamental that servers provide the corresponding security modes.
Hence, it is essential to assess both the strongest (maximum security level that a client can enforce) and the weakest security mode (minimum security level) servers offer.

\begin{figure}[t]
  \centering
  \includegraphics[width=\linewidth]{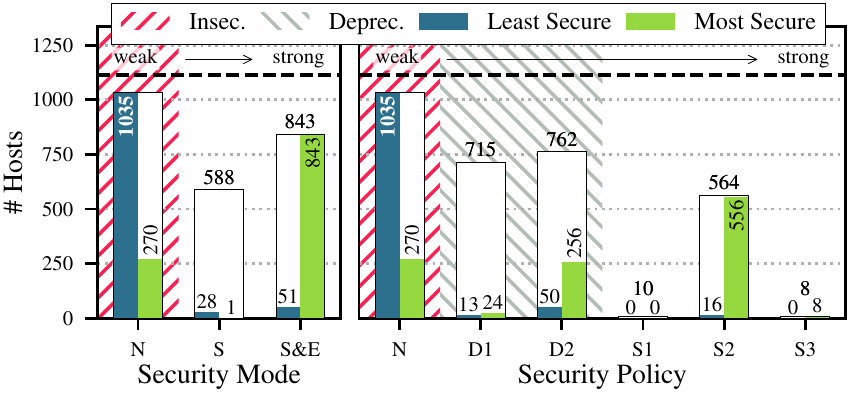}
  \vspace{-2em}
  \caption{
  Number of hosts providing security modes (\texttt{None}~(N), \texttt{Sign}~(S), \texttt{SignAndEncrypt}~(S\&E)) and policies~(cf.\ Table~\ref{tab:policies}) as well as number of hosts offering these as their least and most secure option (dashed line = all hosts).
  }
  \vspace{-1.5em}
  \label{fig:security_modes_policies}
\end{figure}

Figure~\ref{fig:security_modes_policies} (left) shows the number of hosts supporting a specific security mode and further marks the number of hosts where a security mode is the least resp.\ most secure mode available, e.g., the security mode \texttt{Sign}~(S) is supported by \Remark{modescntSignserverslast}~hosts but is the least secure mode available on only \Remark{modescntSignLSMserverslast}~hosts and the most secure mode on only one host, indicating that most of the hosts also support the security modes \texttt{None}~(N) and \texttt{SignAndEncrypt}~(S\&E).
Overall, \Remark{modescntsesMSMserverslast} servers~(\Remark{modespctsesMSMserverslast}\%) follow the recommendation and provide support for at least one of the security modes \texttt{SignAndEncrypt} or \texttt{Sign}, i.e., authenticated communication~(visible in Figure~\ref{fig:security_modes_policies} by summing up hosts supporting one of these modes as their most secure option).
However, \Remark{modescntNoneMSMserverslast} servers~(\Remark{modespctNoneMSMserverslast}\%) only support the security mode \texttt{None} and, therefore, fail to enable secure communication~\cite{opcua-recommendation} rendering the security benefits of OPC~UA inaccessible.

\textbf{Security Policies:}
While message security modes define whether communication security is enabled, security policies define the selected cryptographic primitives~(cf.\ Section~\ref{sec:background}).
Here, two out of five specified security policies other than \texttt{None} are marked as deprecated due to the use of SHA-1~(\texttt{Basic128Rsa15}~(D1), \texttt{Basic256}~(D2); cf.\ Table~\ref{tab:policies}).
Therefore, these policies should not be supported whenever the use of stronger policies is technically possible~\cite{opcua-recommendation}.

Figure~\ref{fig:security_modes_policies} (right) details the number of servers offering security policies and the number of servers providing a specific policy as their least and most secure option.
While the security policy \texttt{None} is only offered in combination with security mode \texttt{None}, the other policies are announced together with the security modes \texttt{Sign} and \texttt{SignAndEncrypt}.
While \Remark{policiescntofferhighserverslast}~servers support one of the policies with a sufficient level of security (S1, S2, S3)~(visible in Figure~\ref{fig:security_modes_policies} by summing up hosts supporting these policies as their most secure option), only \Remark{policiescntenforcehighserverslast} enforce the use of these policies, i.e., do not provide a less secure alternative.
In contrast, \Remark{policiescntofferdeprecatedserverslast}~hosts still support \texttt{SHA-1}-based policies that have been deprecated in 2017~(D1, D2).
A subset of \Remark{policiescntenforcedeprecatedserverslast} servers support these policies as their most secure option and thus lack an adequate security level for connecting clients. %

To summarize, while only \Remark{policiespctenforcehighserverslast}\% of all servers enforce the use of strong cryptographic policies, \Remark{policiespctofferdeprecatedserverslast}\% of all servers still support deprecated and insecure policies.
Despite using OPC~UA, a principally secure protocol, these numbers indicate that the security configurations are not updated as fast as basic cryptographic primitives lose their security guarantees.
Given the long lifetime of industrial components~\cite{nist-icsrecommendations-2015}, exposing such devices to the Internet and not updating their configuration is (potentially) dangerous.

\textit{\textbf{Takeaway:}
Already for elementary security settings, we uncover \Remark{takeawaycntnonelast} servers (\Remark{takeawaypctnonelast}\%) offering no security at all and \Remark{takeawaycntdeprecatedlast} servers (\Remark{takeawaypctdeprecatedlast}\%) supporting only deprecated cryptographic ciphers.
Summing up, \Remark{takeawaypctnonedeprecatedlast}\% of the servers do not fully utilize the security benefits of OPC~UA, i.e., do not allow clients to connect securely according to today's standards.
}

\subsection{Actually Used Security Parameters}
\begin{figure}[t]
    \centering
    \includegraphics[width=\linewidth]{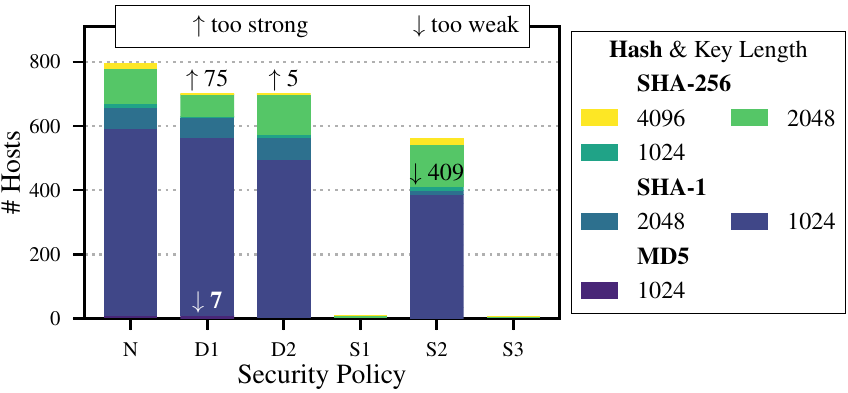}
    \vspace{-2em}
    \caption{Number of servers sending a certificate to implement announced policies (separated by signature hash function and keylength).
    }
    \vspace{-1.5em}
    \label{fig:certificate_quality}
\end{figure}

\begin{figure*}[t]
    \minipage[t]{0.25\textwidth}
        \includegraphics[width=\textwidth]{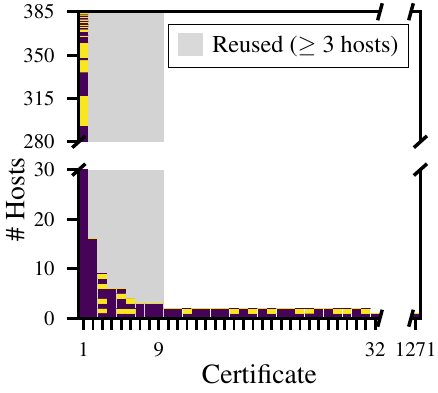}
    	\vspace{-2.2em}
        \caption{Number of hosts using the same certificate to authenticate (alternating colors indicate distinct autonomous systems).}
        \label{fig:reuse}
    \endminipage{}\hfill
    \minipage[t]{0.45\textwidth}
        \includegraphics[width=\textwidth]{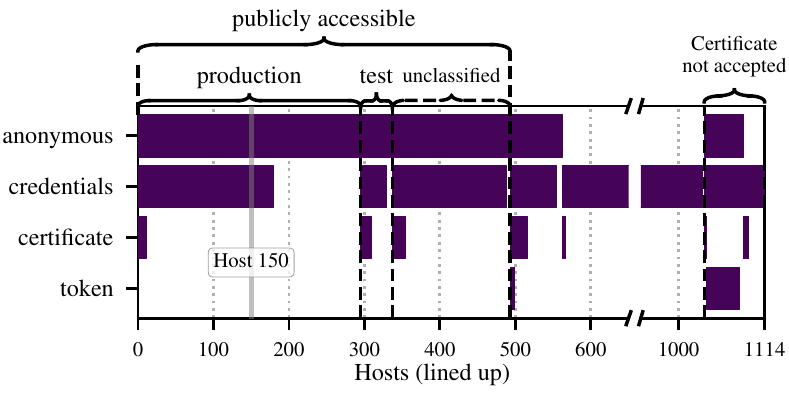}
    	\vspace{-2.2em}
        \caption{All hosts lined up marking offered authentication methods, accessible hosts and our classification results, e.g., Host~150 was accessible, classified as production system and provided anon.\ and cred.\ as auth.\ types.}
        \label{fig:auth}
    \endminipage{}\hfill
    \minipage[t]{0.25\textwidth}
        \includegraphics[width=\textwidth]{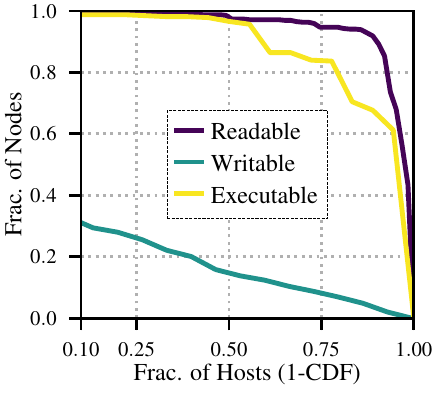}
    	\vspace{-2.2em}
        \caption{Fraction of read- and writeable nodes as well as executable functions on all \Remark{servercntaccessiblelast}~publicly accessible hosts.}
        \label{fig:access}
    \endminipage{}
    \vspace{-0.7em}
\end{figure*}

While security policies define mandatory cryptographic primitives, so far, it was an open question whether OPC~UA servers implement the security parameters determined in their announced policies.
To study the effective implementation of the security policies, we analyze all servers' certificates~(\Remark{certspctselfsignedlast}\% self-signed, 2 CA signed), focussing on their used cryptographic algorithms and key lengths, for conformance with the announced security policies.

In Figure~\ref{fig:certificate_quality}, we present for each security policy the number of received certificates that servers delivered, highlighting the cryptographic hash function and key length.
Most notably, out of the \Remark{policiescntBasic256Sha256serverslast}~servers offering the recommended \texttt{Basic256Sha256}~(S2) policy, \Remark{certscntbasic256sha256nomatch}~servers provide certificates that do not match the specified security parameters, i.e., using \texttt{MD5} or \texttt{SHA-1} and/or a too-short key, weakening the built-in security of OPC~UA.
Contrary, out of the \Remark{policiescntBasic128Rsa15serverslast}~servers announcing the weakest and deprecated security policy~(\texttt{Basic128Rsa15}~(D1)), \Remark{certscntbasic128rsa15toostrong}~servers provide certificates using cryptographic primitives being too ``strong''.
In most cases, these certificates use \texttt{SHA-256} instead of \texttt{SHA-1}~(as required by D1) as signature hash function.
Although too strong primitives do not weaken the security, the specification does not allow such behavior, i.e., such settings potentially hinder clients from connecting.

In general, OPC~UA implementations do not check compliance to security policies or do not sufficiently alert operators when inappropriate certificates are provided.
The use of weaker primitives than defined by the policy causes the gained communication security to be weaker than expected, rendering the security benefits advertised by the security policies ineffective.
Contrary, certificates using too strong primitives nullify the alleged compatibility with legacy clients, which might not be able to operate with these primitives~\cite{hummen-tailoring-2013, hiller-tailoring-2019}.
In case the certificate includes a longer key than the client is able to handle, e.g., due to memory limitations, the client is not able to connect.
However, when the certificate is created using a hash function that the client does not support, the client  might abort the connection or continues without verifying the certificate.

\textit{\textbf{Takeaway:}
Out of the \Remark{modescntsesMSMserverslast} servers, which in \emph{theory} provide sufficient security, \Remark{policiespctofferdeprecatedserverslast}\% actually realize a weaker security level in \emph{practice} than specified, e.g., due to the use of \texttt{SHA-1}.
In combination with the \Remark{policiescntenforcedeprecatedserverslast}~servers already announcing deprecated security options, this leads to \Remark{takeawaypctdeprecatedtooweaklast}\% of servers with deprecated configurations.
Adding the \Remark{takeawaypctnonelast}\% of servers offering no security, \Remark{takeawaypctnonedeprecatedtooweaklast}\% of found servers are affected by configuration deficits.
}

\subsection{Secrets Not Meant to be Shared}
\label{sec:results:certificate-reuse}

Apart from the correct configuration and use of security parameters, the handling of secrets is also crucial for the security of Internet-facing deployments.
Most important, cryptographic secrets must remain private to avert impersonation and eavesdropping~\cite{nist-privatekey-2019}.

To assess whether operators adhere to foundational security best-practices, we check whether different servers authenticate using the same security certificate.
Figure~\ref{fig:reuse} lines up all received certificates and shows the number of hosts announcing it to authenticate.
Notably, we encounter \Remark{certscntreusedlast} certificates of which each is deployed on at least three devices~(to account for devices changing their IP address during our measurements, highlighted in Figure~\ref{fig:reuse}), most likely deployed at different operators.
Although the OPC~UA protocol does not allow the client to verify that the certificate's private key is indeed installed on the server whenever the client certificate is rejected~(due to an ambiguous error response), we still assume so as it is a fundamental requirement for the secure channel handshake~(needed for the decryption of the request)~\cite{opcua-services-2017}.

From available meta-information, we can derive that certificate reuse concerns systems encompassing automation systems, amongst others, for energy technology and parking guidance.
In one extreme case, we found a single certificate issued by a manufacturer of industrial control systems~(as per the subject field) deployed on \Remark{certcntreusebachmannlast} hosts in \Remark{certcntreusebachmannlastuniqueases} different autonomous systems.
Two additional certificates of the same manufacturer show the same practice~(on \Remark{certcntreuseatvise1last} resp. \Remark{certcntreuseatvise2last} devices in \Remark{certcntreuseatvise1lastuniqueases} resp. \Remark{certcntreuseatvise2lastuniqueases} autonomous systems).
We informed the manufacturer about our observations at the end of April~2020, whereupon the manufacturer claimed that distributors and/or operators do not read or understand the product manual which emphasizes the risks of certificate reuse.
As a countermeasure, the manufacturer targeted to sensitize its distributors w.r.t.\ security and sent security information to all it's customers in early June~2020.
However, even three months after the manufacturer sent the security information, we were not able to see any differences in the configuration strategy of the affected hosts.

From the discussion with the manufacturer we can, furthermore, assume that these devices indeed are installed at different operators since a distributor is responsible for the configuration of the devices and sells it to the operators.
Even if not installed at different operators, the usage of the same certificate and key material over hundreds of devices installed in the field, connected to many different ASs, increases the number of attack vectors significantly.

Next to proper handling, the correct creation of secrets is also important for secure deployments, i.e., cryptographic keys often need to be chosen uniformly at random~\cite{heninger-mining-2012}.
While the handling of some secrets is questionable, we have not found any evidence of key material that is subject to insufficient randomness by pairwise checking the keys of all received certificates for shared primes.

\textit{\textbf{Takeaway:} Numerous OPC~UA systems disregard fundamental security recommendations and reuse security certificates across devices, making these susceptible to impersonation and eavesdropping.
As this only affects \Remark{takeawaycntnonedeprecatedtooweakreusedlastadd} devices otherwise configured securely, it has no significant effect on our previous assessment of \Remark{takeawaypctnonedeprecatedtooweakreusedlast}\% of servers being configured deficiently.
}

\subsection{Unprotected OPC UA Address Spaces}
\label{sec:results:address-space}

Access control to protect the information in the address space is an essential part of the OPC~UA security concept, consisting of two steps (cf.\ Section~\ref{sec:background}):
certificate validation during the secure channel establishment, and authorization during the session establishment.
The latter especially means that anonymous access should be disabled, requiring connecting clients to authenticate~\cite{opcua-recommendation,siemens-sinumerik-2018}.

\textbf{Available Authorization Methods:}
Figure~\ref{fig:auth} shows all found OPC~UA servers and details the resp.\ available authentication options\footnote{We further detail on available authentication methods in Appendix~\ref{sec:access:appendix}.}.
We sort servers by the authentication options they support and further separate (to the right) servers that abort the creation of a secure channel rejecting our self-signed certificate (cf.\ Section~\ref{sec:scanning_dataset}).
Out of the \Remark{servercntsecurechanoklast} servers that allow anyone to establish a secure channel, \Remark{servercntsecurechanokanonymouslast}~servers (\Remark{serverpctsecurechanokanonymouslast}\% of all Internet-reachable OPC~UA systems) also offer anonymous access.
Notably, this number encompasses \Remark{takeawaycntpropersecuritybutaccessiblelast}~servers that otherwise force clients to communicate securely.

\textbf{System Classification:}
To evaluate whether systems with insufficient access control indeed are production systems, we access all servers that offer anonymous access not rejecting our session request (\Remark{servercntcertokaccessiblelast} servers, cf. Figure~\ref{fig:auth}), e.g., due to invalid configurations.
To not interfere with any process, we never change the system's state, e.g., by write operations or function executions.
Further, we handle the received data responsibly and reach out to operators whenever we are able to identify them to inform them about their openly accessible system~(cf.\ Appendix~\ref{sec:ethics}).

We heuristically classify accessible OPC~UA systems into production or test systems by analyzing the supported namespaces  (cf.\ Section~\ref{sec:background}).
Although we cannot label \Remark{servercntaccessibleonlydefaultlast} systems (standard namespace only), our approach classifies \Remark{servercntaccessibleproductivelast} production systems based on namespaces relating to industrial manufacturers or standards, e.g., \mbox{IEC 61131-3}~\cite{john-iec-2010}.
Likewise, we categorize \Remark{servercntaccessibletestlast} systems as test systems as they use namespaces of example applications, e.g.,~\cite{freeopcua-examples}.

Overall, our classification points out that a significant ratio of insecure configurations can be linked to production systems allowing control by unauthorized users without any countermeasures.

\textbf{Address Space Access Control:}
To assess the severity of allowing anonymous access, we also analyze the access rights of the anonymous user, i.e., whether servers allow clients reading or writing nodes and executing functions anonymously.

Figure~\ref{fig:access} shows to which extent a share of hosts enables the anonymous user to read or write nodes and execute functions.
\Remark{accesspctrx}\% of all servers allow clients to read more than \Remark{accesspctry}\% of available nodes anonymously, e.g., variables called \texttt{m3InflowPerHour}, indicating that attackers can monitor the device's behavior.
Manual examination of the data the readable nodes contain allowed us to identify a few systems such as parking guidance systems including license plate and video surveillance data.
Further, \Remark{accesspctwx}\% of hosts allow anonymous writes to >\Remark{accesspctwy}\% of their nodes, e.g., \texttt{rSetFillLevel}, enabling attackers to inject various data into the OPC~UA appliance.
\Remark{accesspctex}\% of the systems enable anonymous users to execute over \Remark{accesspctey}\% of functions provided, allowing to change the server configuration, e.g., \texttt{AddEndpoint}.
Based on our judgment, none of the function names suggest that the execution of the function would directly alter the physical state of a machine.
Furthermore, there might be other parts of the production deployment which are under access control.
Still, we find it risky that at least a part of production systems are accessible anonymously enabling attackers to read and write values as well as to execute various functions.

\textit{\textbf{Takeaway:}
\Remark{serverpctaccessiblelast}\% of OPC~UA systems, many classified as production systems, do not realize access control and thus allow anyone to read and write data as well as execute functions.
This issue concerns \Remark{takeawaycntnoneaccessiblelastadd}~servers implementing communication security increasing the share of servers leaving out security opportunities from \Remark{takeawaypctnonelast}\% to \Remark{takeawaypctnoneaccessiblelast}\%.
In total, \Remark{serversinsecurepct}\% of all OPC~UA servers show configuration deficits.
}

\subsection{A Lack of Longitudinal Improvements}
To assess whether the security of OPC~UA configurations improves over time, we
(i)~perform analyzes on weekly data and
(ii)~analyze the distribution of security certificates using included time information covering time beyond the seven months of our measurements.

During seven months~(February to August 2020), we are unable to detect any significant change in the fraction of deficiently configured systems~(avg: \Remark{serverspctdeficitavg}\%, std: \Remark{serverspctdeficitstd}\%, min: \Remark{serverspctdeficitmin}\%, max: \Remark{serverspctdeficitmax}\%).
Still, in 84 cases, we detect certificate renewals on servers with static IP addresses and can investigate whether operators change certificates as part of software updates or use this opportunity to switch to secure ciphers.
In nine cases, we observed a simultaneous software update~(as per OPC~UA's \texttt{SoftwareVersion} field). %
While all renewed certificates were self-signed and valid, only in seven cases security increased by replacing \texttt{SHA-1}. %
Surprisingly, one certificate renewal even resulted in a downgrade from \texttt{SHA-256} to \texttt{SHA-1}.

Analyzing the combined \Remark{certscntall} certificates retrieved over all measurements, we observe that \Remark{certscntsha1issuedafter2017}~(\Remark{certspctsha1issuedafter2017}\%) \texttt{SHA-1} certificates were generated after the deprecation of the corresponding security policies in 2017~(\texttt{NotBefore} field), with \Remark{certscntsha1issuedsince2019} \texttt{SHA-1} certificates still being created and deployed since the beginning of 2019.

Furthermore, we observe continuing deployments of devices from the same manufacturer where multiple devices use the identical certificate (cf.\ Section~\ref{sec:results:certificate-reuse}), noticeably increasing from \Remark{certcntreusebachmannmin} devices on \Remark{certcntreusebachmannmindate} to \Remark{certcntreusebachmannmax} devices on \Remark{certcntreusebachmannmaxdate} indicating that the distributor still installs devices by copying certificates.

\textit{\textbf{Takeaway:}
Weekly scans over seven months and the analysis of certificates generated in a longer timespan show that the security of OPC~UA deployments did not improve over time, e.g., by certificate exchange.
We even observe continuous insecure deployments of systems relying on deprecated ciphers and/or sharing the same certificate.
}

\section{Conclusion}

OPC~UA is the prime candidate to realize standardized and secure industrial communication for an increasing industrial digitalization~\cite{roepert-opcuaassessment-2020}.
In this paper, we study whether the general security of OPC~UA's design~\cite{bsi-opcua-analysis} leads to secure Internet-facing \emph{deployments}.

Using Internet-wide active measurements, we show that \Remark{serversinsecurepct}\% of all \Remark{modescntallserverslast} Internet-reachable OPC~UA deployments are configured deficiently:
First, we reveal servers completely disabling communication security (\Remark{modespctNoneMSMserverslast}\%) or relying on deprecated cryptographic primitives (\Remark{policiespctenforcedeprecatedserverslast}\%) such as \texttt{SHA-1}.
Second, we discover the incorrect application of theoretically secure configurations on additional \Remark{takeawaypctnonedeprecatedtooweaklastadd}\% of systems.
Partly, these systems are also affected by a systematic reuse of security-critical certificates on hundreds of systems across various autonomous systems.
Finally, we find that \Remark{serverpctcertokaccessiblelast}\% of all servers allow unauthenticated users to read and write values from industrial devices and even execute system functionality.

Our analysis in this paper uncovers configuration deficits of OPC~UA devices reachable via IPv4.
It might be possible that various OPC~UA devices are connected via IPv6 only and therefore are not captured in our scans.
We do not anticipate that these devices are configured more securely, but leave this analysis for future research.

To conclude, our results underpin that secure protocols, in general, are no guarantee for secure deployments if they need to be configured correctly following regularly updated guidelines that account for basic primitives losing their security promises.
Thus, we are strongly convinced that it is imperative to reduce configuration complexity in security protocols and demand secure defaults for all configuration options, eventually transitioning from security by design to security by default.
Given the overall still comparatively small number of Internet-connected OPC~UA appliances at the beginning of the uptake of the fourth industrial revolution~\cite{lasi-industry40-2014, pennekamp-towardsiop-2019}, now is the perfect time to rethink and reinforce the security configuration of OPC~UA deployments and thus eventually realize secure industrial Internet-wide communication.

\begin{acks}
\small
We would like to thank the anonymous reviewers and our shepherd Mirja Kühlewind for their fruitful comments.
This work is funded by the \grantsponsor{sponsor_dfg}{Deutsche Forschungsgemeinschaft (DFG, German Research Foundation)}{https://www.dfg.de/en/index.jsp} under Germany’s Excellence Strategy --- EXC-2023 Internet of Production --- \grantnum{sponsor_dfg}{390621612}.
\end{acks}

\bibliographystyle{ACM-Reference-Format}
\bibliography{reference}

\appendix

\section{Ethics}
\label{sec:ethics}

While not involving human subjects, our research still warrants ethical considerations, as measurements of industrial systems could have unintended implications, for example, concerning information security, privacy, or safety.

During the design, execution, and analysis of our research, we thus follow basic ethical research guidelines~\cite{dittrich_menlo-report_2012} as well as practices and procedures imposed by our institutions.
Most importantly, we carefully handle all collected data (cf.\ Appendix~\ref{subsec:ethics:handling}) and further adhere to standard measurement guidelines~\cite{durumeric-zmap-2013} to reduce the impact of our measurements (cf.\ Appendix~\ref{subsec:ethics:impact}).

\begin{figure*}[th!]
  \centering

  \begin{subfigure}[b]{\columnwidth}
    \includegraphics[width=\linewidth]{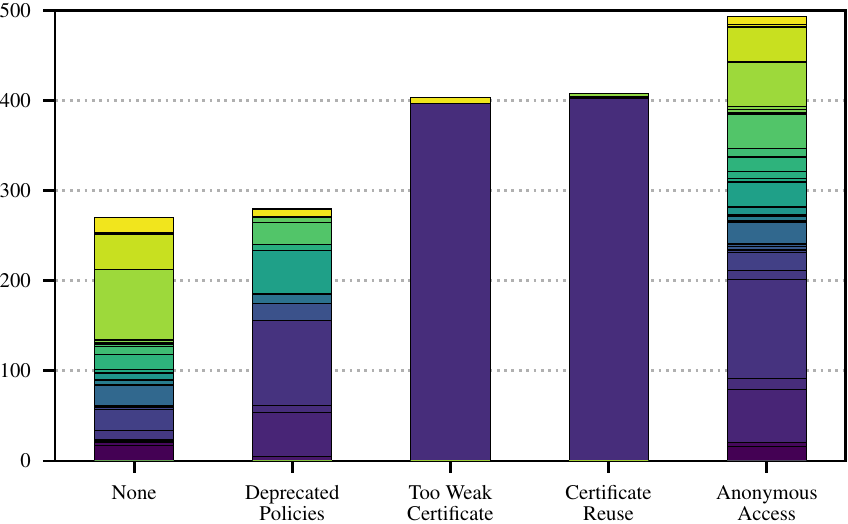}
    \caption{Split by manufacturers~(indicated by different colors)}
    \label{fig:appendix:classes:manufacturer}
  \end{subfigure}
  \hfill %
  \begin{subfigure}[b]{\columnwidth}
    \includegraphics[width=\linewidth]{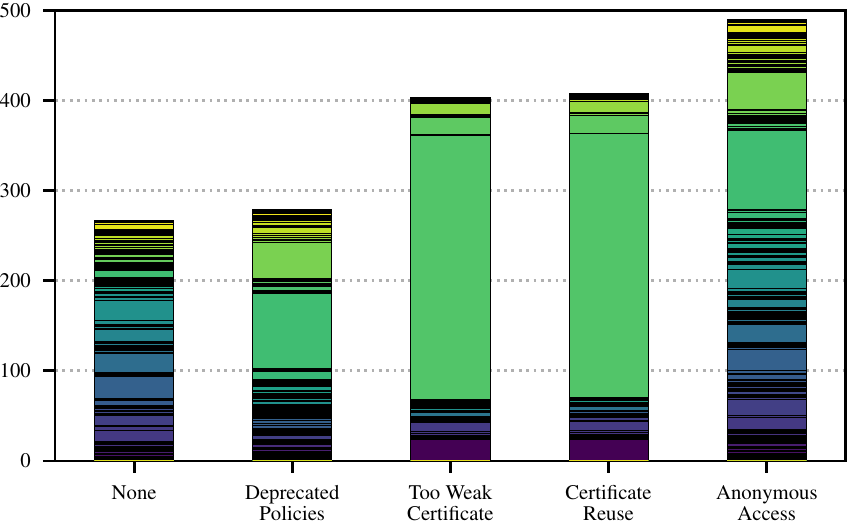}
    \caption{Split by autonomous systems~(indicated by different colors)}
    \label{fig:appendix:classes:as}
  \end{subfigure}
  \vspace{-0.5em}
  \caption{
  Number of hosts only providing security mode and policy \texttt{None}, only deprecated policies (D1, D2, cf.\ Table~\ref{tab:policies}), too weak certificates, affected by certificate reuse, and allow anonymous access.
  }
  \label{fig:appendix:classes}
\end{figure*}

\subsection{Responsible Handling of Data}
\label{subsec:ethics:handling}

As part of our measurements, we only request publicly available data from OPC~UA servers, i.e., data for which no authorization is required.
Thereby, we never bypass any security mechanisms and never alter the state of a server, i.e., we never write to any variable in an OPC~UA server's address space nor execute any functions.
Still, our dataset might contain very sensitive data for those servers that do not implement or do not correctly configure access control for their address space (cf.\ Section~\ref{sec:results:address-space}).
Consequently, we store all data solely on secured systems and exclude collected payload data from our dataset release.
Hence, our results on address space access control in Section~\ref{sec:results:address-space} cannot be independently reproduced by others.
Still, to protect potentially sensitive data, we consider this decision an appropriate trade-off.
To prevent attackers using our dataset to find insecure OPC~UA deployments, we further replace IP addresses and autonomous system IDs by consecutive numbers as well as blacken fields in certificates containing equivalent address information~(e.g., FQDNs).

\textbf{Identifying Operators:}
Besides using the collected data for our research, especially for the classification of systems and analysis of access control (cf.\ Section~\ref{sec:results:address-space}), we analyzed the collected data to identify server operators to inform them about their accessible systems where possible.
To this end, we automatically searched the address space for nodes containing email addresses and, furthermore, invested manual effort to identify the operators of additional systems (where we could not find contact data in the address space).

\textbf{Operator Feedback:}
Overall, out of \Remark{serversaccessiblecnt} systems that provide unprotected access to potentially sensitive data and functionality, we were able to retrieve contact information for 50 systems, including systems for water sewerage, parking lot management, and hotel management.
Subsequently, we reached out to the operators via email (and in one case by phone) to inform them about potential security problems and provided them with pointers to information on how to correctly secure their systems to prevent any potential harm caused by attackers, such as executing system functions as well as stealing or changing data.
We received only two replies to our contact attempts, one promising to forward our information notice to the responsible IT department and one asking for security advice.
At the time of writing (four months after our initial contact attempts), both systems, as well as all but three of the other systems for which we could derive contact information, are still online.
However, the device where the operator asked for security advice now implements access control.

\textbf{Manufacturer Feedback:}
Likewise, we reached out to the manufacturer listed in the subject field of (i) a security certificate used identically on more than 350 hosts as well as (ii) two additional certificates used on fewer hosts (cf.\ Section~\ref{sec:results:certificate-reuse}).
The manufacturer claimed that the issue of certificate reuse likely results from customers/distributors copying system images and/or configuration files containing automatically generated security certificates.
As a consequence, the manufacturer reached out to its customers to sensitize them for the risks resulting from copying security certificates in June~2020.
However, three months after the information was sent, we do not observe any decline in systems using these certificates in our measurements.
Instead, our data suggests that new systems using one of these certificates continue to get deployed as we observe an increase of \Remark{servercntbachmannaddsincemay5} devices since their first reply and \Remark{servercntbachmannweeklyaddlast} devices in the week between our latest measurements.

\subsection{Reducing Impact of Measurements}
\label{subsec:ethics:impact}

While related work claims that measurements with \texttt{zmap} not necessarily impact industrial devices~\cite{coffey-scanninganalysis-2018}, we still set out to minimize the implication of our weekly OPC~UA active Internet measurements to the largest extent possible by following well-established Internet measurement guidelines~\cite{durumeric-zmap-2013}.

\textbf{Restricted Measurements:}
We closely coordinate our measurement study with RWTH Aachen University's Network Operation Center to reduce the impact on our Internet uplink and the Internet as a whole and to handle potential inquiries or abuse requests as fast as possible.
Additionally, we exclude systems from our measurements that requested so. %

\textbf{Contact Information:}
To prominently display the intent of our scans, we provide rDNS records for the IP address used for scanning and provide contact information both in the certificate and the \texttt{ApplicationName} field of our client.
Furthermore, we provide a website at the IP address used for scanning with detailed explanations on the scope and purpose of our research.
Additionally, to allow for exclusion in future scans, we list opt-out instructions on the website.
Based on such requests, we exclude 5.79\,M IP addresses (0.13\% of the IPv4 address space) from our measurements.

\textbf{Measurement Load:}
To not overload any autonomous system, we spread our scans over a timeframe of approximately 24~hours and rely on \texttt{zmap}'s address randomization.
More importantly, to not overload potentially resource-constrained industrial devices during our traversal of their OPC~UA address space, we instruct our scanner module to wait \SI{500}{\milli\second} between subsequent requests to one server.
We further set a scanning time (\SI{60}{\minute}) and outgoing traffic (\SI{50}{\mega\byte}) limit per host, forcing our scanner to disconnect whenever the limit exceeds.
On average, our scanner was connected to an OPC~UA server for \SI{\Remark{measurementtimeavglast}}{\second} (std: \SI{\Remark{measurementtimestdlast}}{\second}, min: \SI{\Remark{measurementtimemsminlast}}{\milli\second}, max: \SI{\Remark{measurementtimemaxlast}}{\second}) causing \SI{\Remark{measurementreadkbavglast}}{\kilo\byte} of outgoing traffic (std: \SI{\Remark{measurementreadmbstdlast}}{\mega\byte}, min: \SI{\Remark{measurementreadbminlast}}{\byte}, max: \SI{\Remark{measurementreadmbmaxlast}}{\mega\byte}).
In some rare cases, our scanner exceeds the set time limit due to side effects caused by parallel scanning of different hosts.

Overall, as OPC~UA does not realize security by default, we consider it essential to know whether modern OPC~UA deployments take advantage of the built-in security features.
To answer this question, we have taken sensible measures to reduce the risks introduced by active Internet measurements of industrial appliances, aiming to influence the security of OPC~UA deployments positively. %

\section{A Closer Look at our Results}
\label{sec:closerlook}
During our research on the security configurations of Internet-facing OPC~UA deployments, we calculated much more extensive statistics than presented in the body of this paper.
Here, we give detailed insight into this data, i.e., we break down our classification of deficiently configured OPC~UA deployments by manufacturers and connecting autonomous systems~(Appendix~\ref{sec:classes:appendix}) as well as elaborate on the access control of OPC~UA devices~(Appendix~\ref{sec:access:appendix}).

\input{appendix_access_control_table.tex_input}

\subsection{Separating Deficits Into Classes}
\label{sec:classes:appendix}
In this paper, we showed that \Remark{serversinsecurepct}\% of Internet-facing OPC~UA deployments are affected by different configuration deficits, i.e., disabling built-in communication security, using deprecated security primitives, disregarding secure policies, systematically reusing private key material, and/or disabling access control.
In this appendix, we shed light on whether devices of specific manufacturers are more affected than devices of other manufacturers~(Appendix~\ref{sec:classes:appendix:manufacturers}) and whether devices in specific autonomous systems are more deficiently configured than devices in other autonomous systems~(Appendix~\ref{sec:classes:appendix:as}).

\subsubsection{Manufacturers}
\label{sec:classes:appendix:manufacturers}

We already described in this paper that some configuration deficits are limited to a few manufacturers, i.e., certificate reuse mainly affects devices of one manufacturer~(cf.\ Section~\ref{sec:results:certificate-reuse}).
In this section, we give insight in whether the other configuration deficits affect the devices built by specific manufacturers more than others.
To determine the device manufacturer of found OPC~UA hosts, we use the same classification as used in the body of this paper~(cf.\ Section~\ref{sec:scanning_dataset}), i.e., we analyzed the \texttt{ApplicationURL} field provided by the servers.

Figure~\ref{fig:appendix:classes:manufacturer} shows for each configuration deficit identified in this paper the number of affected devices and illustrates the distribution over device manufacturers of the hosts.
Indeed, some large fractions of devices built by specific manufacturers are affected by the same configuration deficits, e.g., enabling anonymous access or only providing security mode and policy \texttt{None}.
In one case, only providing security mode and policy \texttt{None} affects all found devices of this manufacturer.
Current product manuals of this manufacturer state that connections relying on this security mode and policy are insecure but do not inform about possible consequences.

The results for too weak security primitives in used certificates and certificate reuse coincides with our results in the main part of the paper, i.e., only a few manufacturers are affected but in one case, these certificates are installed on a large number of devices.

\subsubsection{Autonomous Systems}
\label{sec:classes:appendix:as}

Figure~\ref{fig:appendix:classes:as} details the same analysis but illustrates the distribution of the devices connected to different autonomous systems.
Regarding hosts with too weak certificates in comparison to the security policy description and reusing certificates, a large fraction is connected via the same autonomous system, an ISP focussed on connecting (I)IoT devices to the Internet.
Except for many devices two other autonomous system relying on deprecated policies and allowing anonymous access at the same time~(both being regional Internet service providers), the devices affected by security configuration deficits are distributed over the Internet as they are located in different autonomous systems.

\balance

\subsection{Discovered Access Control Configurations}
\label{sec:access:appendix}

In Section~\ref{sec:results:address-space} we showed that a large number of OPC~UA devices are accessible without any authentication~(offering anonymous access) and elucidated which authentication types are how frequently used.

Table~\ref{tab:accesscontrol} details the distribution of access control settings of Internet-facing OPC~UA servers. On the one hand, \Remark{servercntunaccessiblelast} servers~(\Remark{serverpctunaccessiblelast}\% of all discovered servers) rejected our access attempts by either rejecting the secure channel channel establishment~(\Remark{serverpctunaccessiblesecchanlast}\%) or denying access based on the authentication method~(\Remark{serverpctunaccessibleauthlast}\%).
Note, that this also includes servers initially offering anonymous access but nevertheless aborting the connection due to a faulty or incomplete endpoint configuration.
On the other hand, \Remark{servercntaccessiblelast} servers~(\Remark{serverpctaccessiblelast}\%) allowed access to their address spaces without any authentication although, in many cases, secure authentication types are offered in parallel.
Out of these \Remark{servercntaccessiblelast}~servers, based on used namespaces in their address space, we classified \Remark{servercntaccessibleproductivelast}~servers as production and \Remark{servercntaccessibletestlast}~servers as test systems~(\Remark{serverpctaccessibleproductivelast}\% resp. \Remark{serverpctaccessibletestlast}\% of all Internet-facing OPC~UA deployments).

\end{document}

%% file: appendix_access_control_table.tex_input
\begin{table*}[tbp]
\resizebox{\textwidth}{!}{%
\begin{tabular}{ccccll|c|c|c|lc|c|c|lc}
\cline{1-4} \cline{7-9} \cline{12-13}
\multicolumn{4}{|c|}{\textbf{Authentication Type}} &
   &
   &
  \multicolumn{3}{c|}{\textbf{Accessible Hosts}} &
   &
   &
  \multicolumn{2}{c|}{\textbf{Unaccessible Hosts}} &
   &
  \textbf{} \\ \cline{1-4} \cline{6-9} \cline{11-13}
\multicolumn{1}{|c|}{\textbf{anon.}} &
  \multicolumn{1}{c|}{\textbf{cred.}} &
  \multicolumn{1}{c|}{\textbf{cert.}} &
  \multicolumn{1}{c|}{\textbf{token}} &
  \multicolumn{1}{l|}{} &
  Class: &
  \textbf{Productive} &
  \textbf{Test} &
  \textbf{Unclassified} &
  \multicolumn{1}{l|}{} &
  Reason: &
  \textbf{Authentication} &
  \textbf{Secure Channel} &
   &
  \textbf{} \\ \cline{1-4} \cline{6-9} \cline{11-13} \cline{15-15} 
\multicolumn{1}{|c|}{$\bullet$} &
  \multicolumn{1}{c|}{} &
  \multicolumn{1}{c|}{} &
  \multicolumn{1}{c|}{} &
   &
   &
  \Remark{servercntaccessibleproductiveanonlast} (\Remark{serverpctaccessibleproductiveanonlast}\%) &
  \Remark{servercntaccessibletestanonlast} (\Remark{serverpctaccessibletestanonlast}\%) &
  \Remark{servercntaccessibleunclassifiedanonlast} (\Remark{serverpctaccessibleunclassifiedanonlast}\%) &
   &
   &
  \Remark{servercntunaccessibleauthanonlast} (\Remark{serverpctunaccessibleauthanonlast}\%) &
  \Remark{servercntunaccessiblesecchananonlast} (\Remark{serverpctunaccessiblesecchananonlast}\%) &
  \multicolumn{1}{l|}{} &
  \multicolumn{1}{c|}{\textbf{\Remark{servercntauthanonlast} (\Remark{serverpctauthanonlast}\%)}} \\
\rowcolor[HTML]{C0C0C0} 
\multicolumn{1}{|c|}{\cellcolor[HTML]{C0C0C0}} &
  \multicolumn{1}{c|}{\cellcolor[HTML]{C0C0C0}$\bullet$} &
  \multicolumn{1}{c|}{\cellcolor[HTML]{C0C0C0}} &
  \multicolumn{1}{c|}{\cellcolor[HTML]{C0C0C0}} &
   &
   &
  \Remark{servercntaccessibleproductivecredlast} (\Remark{serverpctaccessibleproductivecredlast}\%) &
  \Remark{servercntaccessibletestcredlast} (\Remark{serverpctaccessibletestcredlast}\%) &
  \Remark{servercntaccessibleunclassifiedcredlast} (\Remark{serverpctaccessibleunclassifiedcredlast}\%) &
   &
   &
  \Remark{servercntunaccessibleauthcredlast} (\Remark{serverpctunaccessibleauthcredlast}\%) &
  \Remark{servercntunaccessiblesecchancredlast} (\Remark{serverpctunaccessiblesecchancredlast}\%) &
  \multicolumn{1}{l|}{\cellcolor[HTML]{C0C0C0}} &
  \multicolumn{1}{c|}{\cellcolor[HTML]{C0C0C0}\textbf{\Remark{servercntauthcredlast} (\Remark{serverpctauthcredlast}\%)}} \\
\multicolumn{1}{|c|}{$\bullet$} &
  \multicolumn{1}{c|}{$\bullet$} &
  \multicolumn{1}{c|}{} &
  \multicolumn{1}{c|}{} &
   &
   &
  \Remark{servercntaccessibleproductiveanoncredlast} (\Remark{serverpctaccessibleproductiveanoncredlast}\%) &
  \Remark{servercntaccessibletestanoncredlast} (\Remark{serverpctaccessibletestanoncredlast}\%) &
  \Remark{servercntaccessibleunclassifiedanoncredlast} (\Remark{serverpctaccessibleunclassifiedanoncredlast}\%) &
   &
   &
  \Remark{servercntunaccessibleauthanoncredlast} (\Remark{serverpctunaccessibleauthanoncredlast}\%) &
  \Remark{servercntunaccessiblesecchananoncredlast} (\Remark{serverpctunaccessiblesecchananoncredlast}\%) &
  \multicolumn{1}{l|}{} &
  \multicolumn{1}{c|}{\textbf{\Remark{servercntauthanoncredlast} (\Remark{serverpctauthanoncredlast}\%)}} \\
\rowcolor[HTML]{C0C0C0} 
\multicolumn{1}{|c|}{\cellcolor[HTML]{C0C0C0}} &
  \multicolumn{1}{c|}{\cellcolor[HTML]{C0C0C0}$\bullet$} &
  \multicolumn{1}{c|}{\cellcolor[HTML]{C0C0C0}$\bullet$} &
  \multicolumn{1}{c|}{\cellcolor[HTML]{C0C0C0}} &
   &
   &
  \Remark{servercntaccessibleproductivecredcertlast} (\Remark{serverpctaccessibleproductivecredcertlast}\%) &
  \Remark{servercntaccessibletestcredcertlast} (\Remark{serverpctaccessibletestcredcertlast}\%) &
  \Remark{servercntaccessibleunclassifiedcredcertlast} (\Remark{serverpctaccessibleunclassifiedcredcertlast}\%) &
   &
   &
  \Remark{servercntunaccessibleauthcredcertlast} (\Remark{serverpctunaccessibleauthcredcertlast}\%) &
  \Remark{servercntunaccessiblesecchancredcertlast} (\Remark{serverpctunaccessiblesecchancredcertlast}\%) &
  \multicolumn{1}{l|}{\cellcolor[HTML]{C0C0C0}} &
  \multicolumn{1}{c|}{\cellcolor[HTML]{C0C0C0}\textbf{\Remark{servercntauthcredcertlast} (\Remark{serverpctauthcredcertlast}\%)}} \\
\multicolumn{1}{|c|}{$\bullet$} &
  \multicolumn{1}{c|}{$\bullet$} &
  \multicolumn{1}{c|}{$\bullet$} &
  \multicolumn{1}{c|}{} &
   &
   &
  \Remark{servercntaccessibleproductiveanoncredcertlast} (\Remark{serverpctaccessibleproductiveanoncredcertlast}\%) &
  \Remark{servercntaccessibletestanoncredcertlast} (\Remark{serverpctaccessibletestanoncredcertlast}\%) &
  \Remark{servercntaccessibleunclassifiedanoncredcertlast} (\Remark{serverpctaccessibleunclassifiedanoncredcertlast}\%) &
   &
   &
  \Remark{servercntunaccessibleauthanoncredcertlast} (\Remark{serverpctunaccessibleauthanoncredcertlast}\%) &
  \Remark{servercntunaccessiblesecchananoncredcertlast} (\Remark{serverpctunaccessiblesecchananoncredcertlast}\%) &
  \multicolumn{1}{l|}{} &
  \multicolumn{1}{c|}{\textbf{\Remark{servercntauthanoncredcertlast} (\Remark{serverpctauthanoncredcertlast}\%)}} \\
\rowcolor[HTML]{C0C0C0} 
\multicolumn{1}{|c|}{\cellcolor[HTML]{C0C0C0}$\bullet$} &
  \multicolumn{1}{c|}{\cellcolor[HTML]{C0C0C0}$\bullet$} &
  \multicolumn{1}{c|}{\cellcolor[HTML]{C0C0C0}} &
  \multicolumn{1}{c|}{\cellcolor[HTML]{C0C0C0}$\bullet$} &
   &
   &
  \Remark{servercntaccessibleproductiveanoncredtokenlast} (\Remark{serverpctaccessibleproductiveanoncredtokenlast}\%) &
  \Remark{servercntaccessibletestanoncredtokenlast} (\Remark{serverpctaccessibletestanoncredtokenlast}\%) &
  \Remark{servercntaccessibleunclassifiedanoncredtokenlast} (\Remark{serverpctaccessibleunclassifiedanoncredtokenlast}\%) &
   &
   &
  \Remark{servercntunaccessibleauthanoncredtokenlast} (\Remark{serverpctunaccessibleauthanoncredtokenlast}\%) &
  \Remark{servercntunaccessiblesecchananoncredtokenlast} (\Remark{serverpctunaccessiblesecchananoncredtokenlast}\%) &
  \multicolumn{1}{l|}{\cellcolor[HTML]{C0C0C0}} &
  \multicolumn{1}{c|}{\cellcolor[HTML]{C0C0C0}\textbf{\Remark{servercntauthanoncredtokenlast} (\Remark{serverpctauthanoncredtokenlast}\%)}} \\
\multicolumn{1}{|c|}{$\bullet$} &
  \multicolumn{1}{c|}{$\bullet$} &
  \multicolumn{1}{c|}{$\bullet$} &
  \multicolumn{1}{c|}{$\bullet$} &
   &
   &
  \Remark{servercntaccessibleproductiveanoncredcerttokenlast} (\Remark{serverpctaccessibleproductiveanoncredcerttokenlast}\%) &
  \Remark{servercntaccessibletestanoncredcerttokenlast} (\Remark{serverpctaccessibletestanoncredcerttokenlast}\%) &
  \Remark{servercntaccessibleunclassifiedanoncredcerttokenlast} (\Remark{serverpctaccessibleunclassifiedanoncredcerttokenlast}\%) &
   &
   &
  \Remark{servercntunaccessibleauthanoncredcerttokenlast} (\Remark{serverpctunaccessibleauthanoncredcerttokenlast}\%) &
  \Remark{servercntunaccessiblesecchananoncredcerttokenlast} (\Remark{serverpctunaccessiblesecchananoncredcerttokenlast}\%) &
  \multicolumn{1}{l|}{} &
  \multicolumn{1}{c|}{\textbf{\Remark{servercntauthanoncredcerttokenlast} (\Remark{serverpctauthanoncredcerttokenlast}\%)}} \\ \cline{1-4} \cline{7-9} \cline{12-13} \cline{15-15} 
 &
   &
   &
   &
   &
   &
  \textbf{\Remark{servercntaccessibleproductivelast} (\Remark{serverpctaccessibleproductivelast}\%)} &
  \textbf{\Remark{servercntaccessibletestlast} (\Remark{serverpctaccessibletestlast}\%)} &
  \textbf{\Remark{servercntaccessibleunclassifiedlast} (\Remark{serverpctaccessibleunclassifiedlast}\%)} &
   &
  \textbf{} &
  \textbf{\Remark{servercntunaccessibleauthlast} (\Remark{serverpctunaccessibleauthlast}\%)} &
  \textbf{\Remark{servercntunaccessiblesecchanlast} (\Remark{serverpctunaccessiblesecchanlast}\%)} &
   &
  \textbf{} \\ \cline{7-9} \cline{12-13}
 &
   &
   &
   &
   &
   &
  \multicolumn{3}{c|}{\textbf{\Remark{servercntaccessiblelast} (\Remark{serverpctaccessiblelast}\%)}} &
   &
   &
  \multicolumn{2}{c|}{\textbf{\Remark{servercntunaccessiblelast} (\Remark{serverpctunaccessiblelast}\%)}} &
   &
  \textbf{} \\ \cline{7-9} \cline{12-13}
\end{tabular}%
}
\vspace{0.4em}
\caption{Number and fraction of hosts supporting different OPC UA authentication types, split by our classification (accessible hosts) and reason for connection reject (unaccessible hosts). Unused combinations of authentication types are omitted.}
\label{tab:accesscontrol}
\end{table*}